\documentclass[fleqn,usenatbib]{mnras}

\usepackage{newtxtext,newtxmath}
\usepackage[T1]{fontenc}

\DeclareRobustCommand{\VAN}[3]{#2}
\let\VANthebibliography\thebibliography
\def\thebibliography{\DeclareRobustCommand{\VAN}[3]{##3}\VANthebibliography}

\usepackage{graphicx}	% Including figure files
\graphicspath{{./plots/}}
\usepackage{amsmath}	% Advanced maths commands
\newcommand{\hGpc}{\,h^{-1}~{\rm Gpc}}

\newcommand{\hMpc}{{\ifmmode{\,h^{-1}{\rm Mpc}}\else{$h^{-1}$Mpc}\fi}}
\newcommand{\hkpc}{{\ifmmode{\,h^{-1}{\rm kpc}}\else{$h^{-1}$kpc}\fi}}
\newcommand{\hMsun}{{\ifmmode{\,h^{-1}{\rm {M_{\odot}}}}\else{$h^{-1}{\rm{M_{\odot}}}$}\fi}}
\newcommand{\Msun}{\,\rm {M_{\odot}}}
\newcommand{\Mstar}{{\ifmmode{\,M_{*}}\else{$M_{*}$}\fi}}
\newcommand{\Mhalo}{{\ifmmode{\,M_{\rm halo}}\else{$M_{\rm halo}$}\fi}}
\newcommand{\ltsima}{$\; \buildrel < \over \sim \;$}
\newcommand{\gtsima}{$\; \buildrel > \over \sim \;$}
\newcommand{\lsim}{\lower.5ex\hbox{\ltsima}}
\newcommand{\gsim}{\lower.5ex\hbox{\gtsima}}

\newcommand{\theth}{\textsc{The Three Hundred}}
\newcommand{\ahf}{\textsc{AHF}}
\newcommand{\caesar}{\textsc{Caesar}}

\newcommand{\gadgetx}{\textsc{Gadget-X}}
\newcommand{\simba}{\textsc{Gizmo-Simba}}
\newcommand{\gadgetmusic}{\textsc{Gadget-MUSIC}}
\newcommand{\galacticus}{\textsc{Galacticus}}
\newcommand{\sag}{\textsc{SAG}}
\newcommand{\sage}{\textsc{SAGE}}
\newcommand{\fable}{\textsc{FABLE}}
\newcommand{\ceagle}{\textsc{C-EAGLE}}

\def\Romeel{\textcolor{black}}

\title[\theth: \simba]{\theth\ project: The \simba\ run}

\author[W. Cui et al.]{Weiguang Cui,$^{1,2}$\thanks{E-mail: weiguang.cui@ed.ac.uk}
Romeel Dave,$^{1}$
Alexander Knebe,$^{2,3,4}$ Elena Rasia,$^{5,6}$ 
Meghan Gray,$^7$ Frazer Pearce,$^7$
\newauthor Chris Power,$^3$ Gustavo Yepes,$^{2,3}$
Dhayaa Anbajagane,$^{8,9}$ Daniel Ceverino,$^{2,3}$ Ana Contreras-Santos,$^{2}$ 
\newauthor Daniel de Andres,$^{2,3}$ Marco De Petris,$^{10}$ Stefano Ettori,$^{11,12}$ Roan Haggar,$^7$ Qingyang Li$^{13}$ Yang Wang,$^{14, 15}$ 
\newauthor Xiaohu Yang,$^{13,16}$ Stefano Borgani,$^{5,6,17}$ Klaus Dolag,$^{18,19}$ Ying Zu,$^{13,16}$ Ulrike Kuchner,$^7$ 
Rodrigo Ca\~{n}as,$^{2,3}$
\newauthor Antonio Ferragamo,$^{10}$ Giulia Gianfagna$^{20}$ 
\\
$^{1}$Institute for Astronomy, University of Edinburgh, Royal Observatory, Edinburgh EH9 3HJ, United Kingdom\\
$^{2}$Departamento de F\'isica Te\'{o}rica, M\'{o}dulo 15, Facultad de Ciencias, Universidad Aut\'{o}noma de Madrid, 28049 Madrid, Spain\\
$^{3}$Centro de Investigaci\'{o}n Avanzada en F\'isica Fundamental (CIAFF), Facultad de Ciencias, Universidad Aut\'{o}noma de Madrid, 28049 Madrid, Spain\\
$^{4}$International Centre for Radio Astronomy Research, University of Western Australia, 35 Stirling Highway, Crawley, Western Australia 6009, Australia\\
$^{5}$IFPU - Institute for Fundamental Physics of the Universe, Via Beirut 2, 34014 Trieste, Italy \\
$^{6}$INAF Osservatorio Astronomico di Trieste, via Tiepolo 11, I-34131, Trieste, Italy\\
$^{7}$School of Physics \& Astronomy, University of Nottingham, Nottingham NG7 2RD, United Kingdom\\
$^8$Department of Astronomy and Astrophysics, University of Chicago, Chicago, IL 60637, USA\\
$^9$Kavli Institute for Cosmological Physics, University of Chicago, Chicago, IL 60637, USA\\
$^{10}$Dipartimento di Fisica, Sapienza Universit\'a di Roma, Piazzale Aldo Moro, 5-00185 Roma, Italy\\
$^{11}$INAF, Osservatorio di Astrofisica e Scienza dello Spazio, via Piero Gobetti 93/3, 40129 Bologna, Italy\\
$^{12}$INFN, Sezione di Bologna, viale Berti Pichat 6/2, 40127 Bologna, Italy \\
$^{13}${Department of Astronomy, School of Physics and Astronomy and Shanghai Key Laboratory for Particle Physics and Cosmology, Jiao Tong University, Shanghai 200240, China}\\
$^{14}$Department of Mathematics and Theories, Peng Cheng Laboratory, No.2, Xingke 1st Street, Nanshan District, Shenzhen 518000, Guangdong Province, P. R. China\\
$^{15}$CSST Science Center for Guangdong-Hong Kong-Macau Great Bay Area, Zhuhai 519082, China\\
$^{16}${Tsung-Dao Lee Institute and Key Laboratory for Particle Physics, Astrophysics and Cosmology, Ministry of Education, Jiao Tong University, Shanghai 200240, China}\\
$^{17}$Department of Physics, University degli Studi di Trieste, Trieste, Italy\\
$^{18}$Universitäts-Sternwarte, Fakultät für Physik, Ludwig-Maximilians-Universität München, Scheinerstr.1, 81679 München, Germany \\
$^{19}$Max-Planck-Institut für Astrophysik, Karl-Schwarzschild-Straße 1, 85741 Garching, Germany\\
$^{20}$INAF, Istituto di Astrofisica e Planetologia Spaziali, via Fosso del Cavaliere 100, 00133 Rome, Italy\\
}

\date{Accepted 2022 May 16. Received 2022 May 16; in original form 2022 February 28}

\pubyear{2022}

\begin{document}
\label{firstpage}
\pagerange{\pageref{firstpage}--\pageref{lastpage}}
\maketitle

\begin{abstract}
We introduce \simba, a new suite of galaxy cluster simulations within \theth\ project. \theth\ consists of zoom re-simulations of 324 clusters with $M_{200}\gtrsim 10^{14.8}M_\odot$ drawn from the MultiDark-Planck $N$-body simulation, run using several hydrodynamic and semi-analytic codes.  The \simba\ suite adds a state-of-the-art galaxy formation model based on the highly successful {\sc Simba} simulation, mildly re-calibrated to match $z=0$ cluster stellar properties. Comparing to \theth\ zooms run with \gadgetx, we find intrinsic differences in the evolution of the stellar and gas mass fractions, BCG ages, and galaxy colour-magnitude diagrams, with \simba\ generally providing a good match to available data at $z \approx 0$.  \simba's unique black hole growth and feedback model yields agreement with the observed BH scaling relations at the intermediate-mass range and predicts a slightly different slope at high masses where few observations currently lie.  \simba\ provides a new and novel platform to elucidate the co-evolution of galaxies, gas, and black holes within the densest cosmic environments.
\end{abstract}


%%%%%%%%%%%%%%%%%%%%%%%%%%%%%%%%%%%%%%%%%%%%%%%%%%

%%%%%%%%%%%%%%%%% BODY OF PAPER %%%%%%%%%%%%%%%%%%

\section{Introduction}

Galaxy clusters are a key class of objects for many astrophysical areas.  On large scales, they are useful for constraining cosmological models via their abundance and evolution. On halo scales, they are interesting sites \Romeel{for} environmental studies of galaxies \Romeel{along} with the evolution of the hot intracluster medium (ICM).  On galactic scales, they are important for studying the oldest galaxy stellar populations and most massive galaxies, \Romeel{along} with the impact of supermassive black holes on galaxies \Romeel{and surrounding gas}. For these reasons, clusters \Romeel{are much investigated} both observationally and theoretically \citep[see][for reviews]{Allen2011,Kravtsov2012,Walker2019}.

Interpreting observations of galaxy clusters from the radio to the X-rays \Romeel{within a structure formation context is} challenging, because clusters contain numerous components interacting over a wide range of scales.  Thus models must capture both the large-scale structure within which clusters grow, while including many small-scale physical processes.  Cosmologically-situated numerical simulations have played an increasingly important role in holistically understanding the physics driving clusters. Unfortunately, clusters are rare objects, so representative cosmological volumes that are able to model all the relevant small-scale physics are extremely challenging computationally.  Many studies have therefore focused on using the zoom simulation technique, where individual clusters are re-simulated with full galaxy formation physics after being extracted from a large (typically dark matter-only) parent simulation.  Zoom simulations must be done one object at a time, but with a sufficiently large sample they can cover the full parameter space of real clusters.

Previous cluster zoom simulations with only dark matter particles, such as Phoenix \citep{Gao2012}, Rhapsody \citep{Wu2013}, \Romeel{and} ZOMG \citep{Borzyszkowski2017} can elucidate the detailed internal structures of the clusters, but cannot directly model the galaxies and gas. Hydrodynamic cluster zoom simulation suites such as Dianoga~\citep{Planelles2013}, MACSIS~\citep{Barnes2017a}, C-EAGLE~\citep{Barnes2017b}, Hydrangea~\citep{Bahe2017}, MUSIC \citep{MUSICI}, and nIFTy~\citep{Sembolini2016}, are able to investigate detailed baryonic properties, and to compare with observations more directly. These are complemented by hydrodynamic simulations that have representative volumes, typically focusing more on the group to poor cluster regime, such as EAGLE~\citep{Schaye2015},  Magneticum~\citep[][]{Dolag2016}, BAHAMAS~\citep{McCarthy2017}, IllustrisTNG~\citep{Pillepich2018}, FABLE~\citep[][which also included zoom regions for galaxy clusters]{Henden2018},   and {\sc Simba}~\citep{Dave2019,Robson2020}.  Thus there is great interest in producing state-of-the-art simulations of clusters, particularly with clusters being a target for numerous forefront observational facilities such as {\it Euclid}, the Dark Energy Survey, eROSITA, Sunyeav-Zeldovich (SZ) telescopes, and the Square Kilometre Array and its precursors.

\theth\ project\footnote{\url{https://the300-project.org/}. {\it the300} is also used for short.} occupies a unique niche among cluster simulation suites. Other suites of cluster simulations have typically focused on a handful of \Romeel{objects, with} zoom regions \Romeel{covering only the cluster and immediate surroundings.  In contrast,} \theth\ re-simulates a mass-complete sample of 324 galaxy clusters extracted from the MultiDark cosmological simulation, using the zoom region that extends out to many virial radii.  The penalty for having so many clusters with large zoom regions is that the numerical resolution is necessarily lower owing to computational limitations.  However, the benefit is that it covers a relatively wide and complete halo mass range, enables larger-scale cosmic web studies around clusters, and provides good statistics along with the ability to investigate rare systems. Furthermore, another interesting feature of \theth\ project is that all these clusters have been run with several different galaxy evolution codes. These include the cosmological hydrodynamic codes \gadgetmusic\ \citep{MUSICI} and \gadgetx\ \citep{Rasia2015}, as well as three different semi-analytical models (SAMs): \galacticus\ \citep{Benson2012}, \sage\ \citep{Croton2016} and \sag\ \citep{Cora2018}. This enables cross-comparisons between models employing different input physics, to better understand the sensitivity to the various physical processes and the robustness of the resulting predictions.  

In this paper, we introduce another set of hydrodynamic runs to \theth\ suite, namely the \simba\ runs \footnote{Note that to distinguish from the {\sc Simba} simulation~\citep{Dave2019} which is a 100$\hMpc$ cosmological hydrodynamic simulation run with the same code, this run for \theth\ clusters is referred to as \simba\ run.}.  This suite uses the {\sc Gizmo} cosmological hydrodynamics code in its Meshless Finite Mass (MFM) solver mode, as opposed to the {\sc Gadget}-based runs which use Smoothed Particle Hydrodynamics (SPH).  It further includes a suite of galaxy formation physics similar to that in the recent {\sc Simba} simulation~\citep{Dave2019} that yields an excellent match to a wide range of galaxy, black hole, and intergalactic medium properties.  
Its novel input physics modules such as torque-limited black hole growth, stably bipolar jet feedback, and on-the-fly dust tracking, make it a valuable addition to the existing \theth\ suite. 

This paper is organised in the following order: we first introduce the \theth\ project in \S\ref{sec:300}. Then, we present the details of the new \simba\ run in \S\ref{sec:simba}. The general comparisons to the other models and observation results are shown in \S\ref{sec:results}. We also include the gas scaling relations in the appendixes which seem less affected. At last, we conclude and discuss our results in \S\ref{sec:conc}.

\section{\theth\ project}
\label{sec:300}
\theth\ project \citep{300Cui2018} is a set of cluster-scale zoom simulations based on a mass-complete sample of 324 most massive galaxy clusters ($M_{\rm vir} \gtrsim 8\times 10^{14}\hMsun$)\footnote{The halo mass is defined as the mass enclosed inside an overdensity of $\delta$ times the critical density of the universe: $\delta = \sim98$ for virial mass at $z=0$ \citep{Bryan1998} and $M_{200, 500}$ is with $\delta = 200, 500$ respectively. Similarly, $R_{500}$ is the radius at which the overdensity $\delta = 500$ is reached.} drawn from the MultiDark simulation (MDPL2, \citealt{Klypin2016}).  MDPL2 assumes cosmological parameters from {\it Planck} \citep{planck2016}, and has a periodic cube of comoving length $1 \hGpc$ containing $3840^3$ DM particles having a mass of $1.5 \times 10^9 \hMsun$ each. Each cluster region was selected to have a comoving radius of $15 \hMpc$ (over $5\times R_{200}$) for re-simulation with different baryonic models: \gadgetmusic~\citep{MUSICI}, \gadgetx~\citep{Rasia2015,Beck2016}, and now \simba~(this work). Additionally, galaxy catalogues in the same cluster regions are extracted from three different SAMs that were run on MDPL2 \citep{Knebe2018}: \sag~\citep{Cora2018}, \sage~\citep{Croton2016}, \Romeel{and} \galacticus~\citep{Benson2012}. 
% The data from all these runs are available by request to \theth\ Steering Committee.

The re-simulation regions are generated with the parallel {\textsc GINNUNGAGAP} code\footnote{\url{https://github.com/ginnungagapgroup/ginnungagap}}: the highest resolution Lagrangian regions share the same mass resolution as the original MDPL2 simulation with gas particles ($M_{\rm gas} = 2.36\times10^8 \hMsun$) split from DM particles. The outside regions are degraded in multiple layers (with a shell thickness of $\sim 4 \hMpc$) with lower mass resolution particles (mass increased by eight times for each layer) that eventually provide the same tidal fields at a much lower computational costs than in the original simulation. 

\Romeel{The afore}mentioned unique features of \theth\ project \Romeel{has enabled} many studies \Romeel{of various} aspects of galaxy clusters \Romeel{to be} carried out. To date, \theth\ has been used in over 30 papers investigating galaxy clusters and their environs. These include \Romeel{studying} the detailed relationship between the central cluster and connecting filaments \citep{300Rost2021, 300Kuchner2020,300Kuchner2021a}, the feeding of the galaxy clusters\citep{300Kuchner2021b,300Kotecha2021}, cluster backsplash galaxies \citep{300Haggar2020,300Knebe2020}, and the virial shock radius \citep{300Baxter2021,300Anbajagane2021b}. The advanced input physics in the hydrodynamic simulations \Romeel{further} allow detailed investigations on cluster properties, such as cluster profiles \citep{300Mostoghiu2019,300Li2020}, substructure and baryon content \citep{300Arthur2019,300Haggar2021,300Mostoghiu2021,300Mostoghiu2021b}, dynamical state and morphologies \citep{300Capalbo2021a,300DeLuca2021,300Capalbo2022,Zhang2021}, ICM (non-)thermalization \citep{300Sayers2021,300Sereno2021}, the fundamental plane \citep{300Diaz-Garcia2021}, the effects of mergers on the BCG properties \citep{300Contreras-Santos2022}, and \Romeel{various} methods for estimating galaxy cluster masses, namely dynamics \citep{300Ansarifard2020,300Li2021,300Li2022}, hydrostatic equilibrium \citep{300Gianfagna2022}, \Romeel{and} machine learning \citep{300DeAndres2022,300DeAndres2022b}. Lastly, comparing to the void/field region runs in this project allows us to study the effect of environment \citep{300Wang2018}; a self-interacting dark matter run \Romeel{was done} that allows constraints on the dark matter cross-section \citep{300Vega-Ferrero2021}; and even chameleon gravity \Romeel{was examined} \citep{300Tamosiunas2021}.  With many more projects in the works, it is valuable to continue to update \theth\ runs with state-of-the-art physical models \Romeel{in order} to expand its range and robustness.
\section{The \simba\ run}
\label{sec:simba}

\subsection{The {\sc Simba} model}

The \simba\ runs of \theth\ are performed with the {\sc Gizmo} code \citep{Hopkins2015} with the state-of-the-art galaxy formation subgrid models following the {\sc Simba} simulation~\citep{Dave2019}. We refer the interested reader to \cite{Dave2019} for full details of all of {\sc Simba}'s features, and here focus on its more unique aspects relevant for clusters.  We also describe our modifications to the {\sc Simba} model parameters utilised for the \simba\ runs of \theth\ clusters, which required re-tuning owing to the lower numerical resolution in \simba\ relative to the original {\sc Simba} simulation.

Owing to the Meshless Finite Mass (MFM) solver implemented in {\sc Gizmo}, gas particles are evolved following an accurate description of shocks and shear flows, without the need for any artificial viscosity. This feature improves the description of shocks and flows with high Mach number, which provides a realistic simulation of outflows and jets.  It also provides improved handling of contact discontinuities relative to SPH.  See \cite{Hopkins2015} for a full discussion of the differences of MFM with respect to other hydrodynamics methods.

Radiative cooling and photon-heating/ionization processes of gas are implemented using the \texttt{Grackle-3.1} library \citep{Smith_2017}, which also accounts for metal cooling with non-equilibrium primordial chemistry treatment. {\sc SIMBA} adopts a spatially-uniform \citep{Haardt2012} ultraviolet background model, accounting for self-shielding on the fly based on the prescription in \cite{Rahmati_2013}. An $\rm H_2$-based star formation model is taken from its predecessor simulation {\sc Mufasa} \citep{Dave2016}, which is calibrated to match the \cite{Schmidt_1959} law. Here $\rm H_2$ is estimated from the local column density and metallicity following the \citet{Krumholz_2009,Krumholz_2011} prescription. Besides requiring that $H_2$ be present, an additional minimum density cut of $n_{\rm H}> 0.1~{\rm cm}^{-3}$ and a minimum metallicity of 0.05 for the $f_{\rm H_2}$ in its formation \citep{Krumholz_2009}, compared to the original {\sc Simba} simulation of $0.1~{\rm cm}^{-3}$ and 0.01, respectively, are required for active star formation.

Star formation-driven galactic winds also shares the same decoupled two-phase model in \textsc{Mufasa}, but the mass loading factor scaling with stellar mass is based on the Feedback in Realistic Environments (FIRE) zoom simulations of \citet{Angles-Alcazar_2017b}:
\begin{equation}
    \eta(M_*) \approx \begin{cases} 9(\frac{M_*}{M_0})^{-0.317} &\text {if $M_* < M_0$} \\
    9(\frac{M_*}{M_0})^{-0.761} & \text {if $M_* \ge M_0$} \end{cases}.
\end{equation}
Here, $M_0 = 2 \times 10^9 \Msun$; this is slightly different than the original {\sc Simba} simulation, as we will motivate later. The ejection velocity is based on scalings from \cite{Muratov_2015} as in {\sc Mufasa}:
\begin{equation}
    v_w = 0.854 \left(\frac{v_{\rm circ}}{200\ {\rm km s^{-1}}}\right)^{0.12} v_{\rm circ} + \Delta v(0.25 R_{\rm vir}),
\end{equation}
where $\Delta v(0.25R_{\rm vir})$ is the velocity corresponding to the potential difference between the launch point and one-quarter of the virial radius. Again, this has been changed relative to {\sc Simba}, who used a normalisation of 1.7;
the normalisation used here in \simba\ is the original one proposed by \cite{Muratov_2015}.  Identically to {\sc Simba}, galaxies are identified with the on-the-fly approximate friends-of-friends (FOF) finder for star, dense gas and BH particles in \cite{Dave2016}, allowing galaxy properties such as $M_*$ \Romeel{to be} computed on-the-fly, with $v_{\rm circ}$ obtained from a scaling based on the \Romeel{observed} baryonic Tully–Fisher relation \citep{McGaugh2012}.

The chemical enrichment model tracks eleven elements (H, He, C, N, O, Ne, Mg, Si, S, Ca, Fe), \Romeel{with metals} from supernovae type Ia \citep{Iwamoto1999} and type II \citep{Nomoto2006}, and Asymptotic Giant Branch (AGB) stars \citep{Oppenheimer_2006}. Furthermore, {\sc Simba} also \Romeel{includes} metal-loaded winds, i.e. metals in the wind particle \Romeel{are enhanced, and correspondingly} subtracted from nearby gas in a kernel-weighted manner.  All this is identical to the original {\sc Simba} model; see \citet{Dave2019} for details.

{\sc Simba} seeds black hole (BH) particles based on the host galaxy stellar mass, $M_* > \gamma_{\rm BH} \times M_{\rm seed}$. If the galaxy meets the aforementioned condition and does not already contain a black hole particle, then the star particle closest to the centre of mass of the galaxy is converted into a black hole particle. For this \simba\ run, we employ $M_{\rm seed} = 10^5 \hMsun$ and $\gamma_{\rm BH} = 3 \times 10^5$, which sets the galaxy stellar mass threshold for seeding BH is $M_* \approx{} 10^{10.5} \hMsun$. This is 10 times higher than original {\sc Simba} model\Romeel{, owing to the lower resolution}. By assuming the dynamical friction is efficient enough to maintain black holes near the host galaxy’s centre (within 4 times the size of the BH kernel, $R_0$, considered for the accretion model), black hole particles are re-positioned to the location of the potential minimum within the FOF host group at each time-step. Furthermore, any two black holes located within $R_0$ are allowed to merge instantaneously if their relative velocity is lower than three times their mutual escape velocity.

The BH accretion follows a dual model: The cold accretion mode is described with a torque-limited accretion model for the cold gas ($T \leq 10^5 K$), driven by disk gravitational instabilities arising from galactic scales down to the accretion disk around the central BH (\citealt{Hopkins_2011}; see also \citealt{Angles-Alcazar_2013, Angles-Alcazar_2015, Angles-Alcazar_2017a}).  Hot gas ($T > 10^5 K$) is accreted based on the Bondi rate \citep{Bondi_1952}. We reduce the Bondi accretion rate to the rate for a $M_{\rm BH} = 10^9 \hMsun$ BH, no matter how big it gets; in {\sc Simba}, this was set to $10^{10}M_\odot$.  The black hole accretion kernel has a distance enclosing 256 \Romeel{baryonic} particles or $R_{0} = 6 \hkpc$ (comoving), whichever is smaller, within which the gas quantities are calculated; the latter was \Romeel{set to} $2\hkpc$ in {\sc Simba}.  The total accretion rate for a given black hole is then the sum of $\dot M_{\rm Torque}$ and $\dot M_{\rm Bondi}$, times an additional constant $1 - \eta$, with the radiative efficiency \Romeel{assumed to be} $\eta = 0.1$ \citep{Dave2019}. This total accretion rate is used to determine the AGN feedback modes detailed in the following paragraph.

There are three different AGN feedback modes: A kinetic subgrid model for both the `radiative mode' and `jet mode', and a mostly kinetic X-ray feedback mode accounting for \Romeel{radiation pressure from} X-rays off the accretion disc broadly following \cite{Choi2012}. The `radiative mode' feedback is turned on when the BH is accreting at a high Eddington ratio ($f_{\rm Edd} \equiv \dot M_{\rm BH} / \dot M_{\rm Edd} > 0.2$). The radiative velocity for wind particles, which is based on ionized gas linewidth observations of X-ray detected AGN from SDSS \citep[see Fig. 8 in][]{Perna2017}, scales as:
\begin{equation}
    v_{w,\ \rm Radiative} = 500+\frac{500}{3} \left(\log_{10} \frac{M_{\rm BH}}{\Msun}+6 \right) \, \rm km \, s^{-1}.
\end{equation}
When the BH's in low Eddington accretion mode, $f_{\rm Edd} < 0.2$, the wind begins to transition into a jet mode, with the velocity scaled with $f_{\rm Edd}$ as follows:
\begin{equation}
    v_{w,\ \rm Jet}= v_{w,\ \rm Radiative} + 15000 \log_{10} \left( \frac{0.2}{f_{\rm Edd}} \right) \, {\rm km \, s^{-1}}.
\end{equation}
Note that the wind in both modes is ejected in the form of purely bipolar outflows, based on the angular momentum of gas and stars within $R_0$.  The wind velocity in the jet mode is capped at $15000 \, \rm km/s$ (as opposed to $7000$~km/s in original {\sc Simba}) when the Eddington rate drops below $f_{\rm Edd} \leq 0.02$.  There is another condition to trigger jet mode -- the minimum BH mass has to be greater than $10^{7.5} \Msun$. X-ray feedback only operates when the `jet mode' AGN feedback is in action, and it further has to meet another two conditions: $M_* > 10^9 \Msun$ and $M_{\rm gas}/M_{\rm baryon} < 0.2$. The $M_*$ condition is raised with respect to the original {\sc Simba} model, owing to the lower resolution.

Another novel feature of the {\sc Simba} simulation, the on-the-fly dust production and destruction model, is also employed in \simba, unchanged.  This model is quite successful in reproducing galaxy dust properties over cosmic time \citep{Li2019}.

\begin{table*}
    \centering
    \begin{tabular}{l|c|c}
        Parameter & Original {\sc Simba} simulation & \simba\ run for the300 clusters\\
        \hline
        $n_{\rm H}$ threshold for SF & 0.13 & 0.1 \\
        Metallicity floor for $f_{\rm H2}$ & 0.001 & 0.05 \\
        $M_0$ in SN mass loading factor & $5.2 \times 10^9 \Msun$ & $2 \times 10^9 \Msun$\\
        Galaxy stellar mass limit for seeding BH & $10^{9.5} \hMsun$ & $10^{10.5} \hMsun$\\
        BH mass for Bondi accretion rate cap & $10^{10} \hMsun$ & $10^9 \hMsun$\\
        BH accretion kernel radius & $2 \hkpc$ & $6 \hkpc$\\ 
        Cap wind velocity limit in jet mode & $7000 \, \rm km/s$ & $15000 \, \rm km/s$ \\
        Galaxy stellar mass limit for X-ray feedback & 0 & $10^9 \Msun$ \\
        Gravitational softening length & $0.5 \hkpc$ minimum &  $5 \hkpc$ fixed \\
        \hline
    \end{tabular}
    \caption{Summary of parameter changes with respect to the original {\sc Simba} simulation.}
    \label{tab:1}
\end{table*}

\subsection{Re-calibration for \theth}

We now summarise the modifications from the original {\sc Simba} simulation, and describe the datasets used for the re-calibration. The set of parameter changes is tabulated in Table~\ref{tab:1}.  To reiterate, this is required because \theth\ simulation has about 10 times worse mass resolution than the {\sc Simba} simulation.  It has been found that the {\sc Simba} model is reasonably well converged towards higher resolutions than in the original {\sc Simba} simulation, but it has so-called weak convergence~\citep{Schaye2015} towards poorer resolution, and must be re-calibrated to achieve an equivalently good match to observations.

We use three observational relations focusing on the stellar component in galaxy clusters to calibrate the model parameters for our \simba\ runs, all at $z\approx 0$:  (C1) total stellar mass fraction within $R_{500}$; (C2) BCG stellar mass -- halo mass relation; and (C3) the satellite galaxy stellar mass function (SSMF) in galaxy clusters. The denotations C1, C2, and C3 are further noted in the three subsection titles in \S\ref{sec:results} where we show comparisons to some calibration datasets.  We note that the calibration was done mostly by trial-and-error using intuition to guide the variations, so should not be regarded as a unique parameter set that achieves agreement.

The calibration was not done over the entire sample, but rather only using a single cluster region \Romeel{at $z=0$} whose largest object has $M_{500}\approx 5\times 10^{14}M_\odot$, chosen to be typical of \theth\ sample\footnote{Ideally, we could calibrate the parameters for each cluster region, then used the mean or median values to rerun the whole cluster data set. However, this requires an infeasible amount of computation time. Furthermore, the observation results have a large scatter, and the sub-grid models are not expected to perfectly match all calibration data. Thus, we only require the calibrated parameters yield results in rough agreement with observations.}. Additionally, all uncontaminated halos (i.e. those that do not contain any low-resolution particles) within that region are considered (which includes about 10 more halos down to $\approx 10^{13}M_\odot$), although typically these do not add much information since the observational constraints tend to get weaker towards lower masses.  Once sufficiently calibrated based on this single region, all the parameters in Table~\ref{tab:1} were frozen and run for all the remaining galaxy clusters.  As such, while the agreements to the datasets used for C1, C2, and C3 are to be considered as tuned and not an intrinsic success of the model, it is also the case that in principle other cluster regions could have shown large variations with respect to the one region used for tuning, particularly at different masses.  Hence the agreements with C1, C2, and C3 over the entire mass range (and to higher redshifts, where applicable) may still be regarded as a modest success.

We now describe the motivation for the individual changes, namely what went wrong with the cluster run when we adopted the original {\sc Simba} parameters.
At this low resolution, we found that the star formation is insufficient \Romeel{within} satellite galaxies, resulting in a much lower satellite stellar mass function (SSMF) compared to observations (C3). The changes for $n_{\rm H}$, Metallicity floor, $M_0$ in SN mass loading factor and galaxy stellar mass limit for X-ray feedback serve to boost the SF in these satellite galaxies. However, as a consequence of these changes, the total stellar mass and BCG stellar mass ended up significantly higher compared to observations (C1 and C2).  Thus we strengthened the AGN jet feedback, which has been shown as the key to quench massive galaxies~\citep[][]{Cui2021}, by increasing the maximum jet speed; this reduces the stellar mass of BCG (and hence total).  Finally, we choose to use a fixed comoving softening length of $5 \hkpc$ for consistency with the other \theth\ runs, as opposed to a variable softening length with a minimum of $0.5 \hkpc$ as in original {\sc Simba}.  The BH accretion kernel maximum radius was commensurately increased in \simba\ owing to the coarser spatial resolution, and the BH seeding stellar mass was increased by an order of magnitude to reflect the order of magnitude poorer particle mass resolution.

\subsection{Detailed differences between \gadgetx\ and \simba}
\begin{table*}
    \caption{The detailed model differences between \gadgetx\ and \simba.}
    \label{tab:model_diff}
    \centering
    \begin{tabular}{@{}l@{\:}c@{\:}c@{}}
    % \begin{tabular}{c|c|c}
      modules  & \gadgetx\ & \simba\ \\
     \hline
     & {\bf Gravity and hydro solvers} & \\
     \hline
     Simulation code    & Gadget-3P & GIZMO (Gadget-3 based) \\
     Gravity solver     & TreePM    & TreePM \\
     Hydro solver       & SPH       & MFM \\
     Kernel for hydro solver    & Wendland C4  & Cubic spline \\
     \hline
     & {\bf baryon models -- gas} & \\
     \hline
     Gas cooling        & Metal cooling table \citep{Wiersma2009} & Grackle-3.1 library \citep{Smith_2017} \\
     UV/X-ray background radiation  & \cite{Haardt2001} & \cite{Haardt2012}\\
     Self-shielding     & No        & \cite{Rahmati_2013} \\
     \hline
     & {\bf baryon models -- star} & \\
     \hline
     Star formation model & \cite{Tornatore2007} & \cite{Dave2016}\\
     Star formation rate    & Gas density and temperature based & $\mathrm{H}_2$-based\\
     Star generations$^a$ & Multiple  &   Single \\
     IMF                & \cite{Chabrier2003}   & \cite{Chabrier2003}\\
     Chemical enrichment model & 11 elements from SN-II, SN-Ia, and AGB stars & The same$^b$ \\
     Stellar feedback   & kinetic feedback \citep{Springel2003} & Two-phase winds with mass loading factor depending on $M_*$\\
     \hline
     & {\bf baryon models -- BH} &\\
     \hline
     BH seeding condition  & $M_{FoF}>8\times10^{11} \hMsun\ \&\ M_*>1.6\times10^{10} \hMsun$ & Galaxy stellar mass $> 3\times 10^{10} \hMsun^c$\\
     BH seed mass  & $ 5\times10^{6} \hMsun$ & $ 10^{5} \hMsun$\\
     BH accretion   &  Bondi accretion  & torque-limited and Bondi accretion models \\
     BH feedback    &  Thermal feedback$^d$ & Kinetic feedback$^e$ + X-ray feedback$^f$\\
     \hline
     \multicolumn{3}{l}{$^a$ Number of stars that per gas particle can spawn.}\\
     \multicolumn{3}{l}{$^b$ Note that the detailed implementations are different, see \cite{Tornatore2007} and  \cite{Dave2016} for more information.}\\
     \multicolumn{3}{l}{$^c$ Note here $M_*$ is the total stellar mass within the FoF halo. And another two conditions have to be met as well: $M_*>0.05 M_{DM}\ \&\ M_{gas}>0.10M_*$}\\
     \multicolumn{3}{l}{$^d$ The feedback energy is coming from both mechanical and radiative modes.}\\
     \multicolumn{3}{l}{$^e$ Different outflow velocities are used to mimic the jet and radiative mode AGNs in observation.}\\
     \multicolumn{3}{l}{$^f$ Thermal or thermal+kinetic feedback is used for the X-ray heating depending on whether the surrounding gas is non-ISM or ISM, respectively.}\\
    \end{tabular}
\end{table*}

As noted in the Introduction, we will focus on the differences between the results from \gadgetx\ and \simba\ runs, given that they are the two hydrodynamic models that match a reasonably wide suite of observations. Though both simulation codes are well presented in a series of literature papers, it is useful to relist their key features again in \autoref{tab:model_diff} for comparison. For more detailed models of \gadgetx, we refer to \cite{300Cui2018} and references therein. Although we give descriptions of the {\sc Simba} model in this \autoref{sec:simba}, interested readers are referred to \cite{Dave2019} and \cite{Dave2016} for further information. Besides these model differences as shown in \autoref{tab:model_diff}, we emphasise that \gadgetx\ is quite successful in reproducing the observed gas properties and relations, while the original {\sc Simba} simulation was primarily tuned to reproduce galaxy stellar properties. Here, we follow {\sc Simba}'s process by calibrating \simba\ according to the observed stellar properties, with no regard to gas properties. Comparing the two runs in different properties will help us to better constrain galaxy formation models. 

\subsection{The \ahf\ halo and \caesar\ galaxy catalogues}
Two object catalogues are generated from the suite of the300 galaxy cluster runs: the \ahf\footnote{\url{http://popia.ft.uam.es/AHF}} \citep{Knollmann2009} halo catalogues and the \caesar\ \footnote{\url{https://github.com/dnarayanan/caesar}} galaxy/FoF halo catalogues.
\ahf\ provides halo and subhalo (and thus galaxy) catalogue generated using a spherical overdensity (SO) algorithm, while \caesar\ also provide halo catalogue generated using a 3-D FoF algorithm along with a matched galaxy catalog using a 6-D FoF. \caesar\ further provides a large range of pre-computed physical and photometric properties (with and without dust extinction) for each object.

In this paper, we use as many (uncontaminated) objects as possible (if not specified) to do the investigation because \theth\ runs have a much larger radius, thus many smaller mass halos besides the central cluster in each region. If not specified, we always use the halo mass defined as $M_{500}$ with the quantities are always calculated within $R_{500}$. Therefore,  halo properties from the \ahf\ catalogue are used to compare with the global clusters properties from observations, while galaxy properties from \caesar\ which has a 6D (in both spatial and velocity field) galaxy finder, is used. We further match the galaxies from \caesar\ to the halos from \ahf, by simply taking all the \caesar\ galaxies within the \ahf\ halo radius ($R_{500}$). The BCG is selected as the most massive \caesar\ galaxy that lies close to the \ahf\ distinct halo centre. As the 6D galaxy finder makes no distinction between central and satellite galaxies, it doesn't matter whether the galaxy is coming from a FoF halo or a SO halo.

To track cluster growth histories, we use the cluster main progenitors determined by the {\sc MERGERTREE} package integrated into the \ahf\ program. The main progenitors are selected based on the matched dark matter particle IDs.

\section{results}
\label{sec:results}

\subsection{Baryon fractions}
We first focus on the total gas and stellar components within the clusters (mostly within $R_{500}$) in this subsection. We will also detail their evolution and show the differences between \gadgetx\ and \simba\ runs.

\subsubsection{C1: The gas and stellar fractions}
\begin{figure*}
	\includegraphics[width=\textwidth]{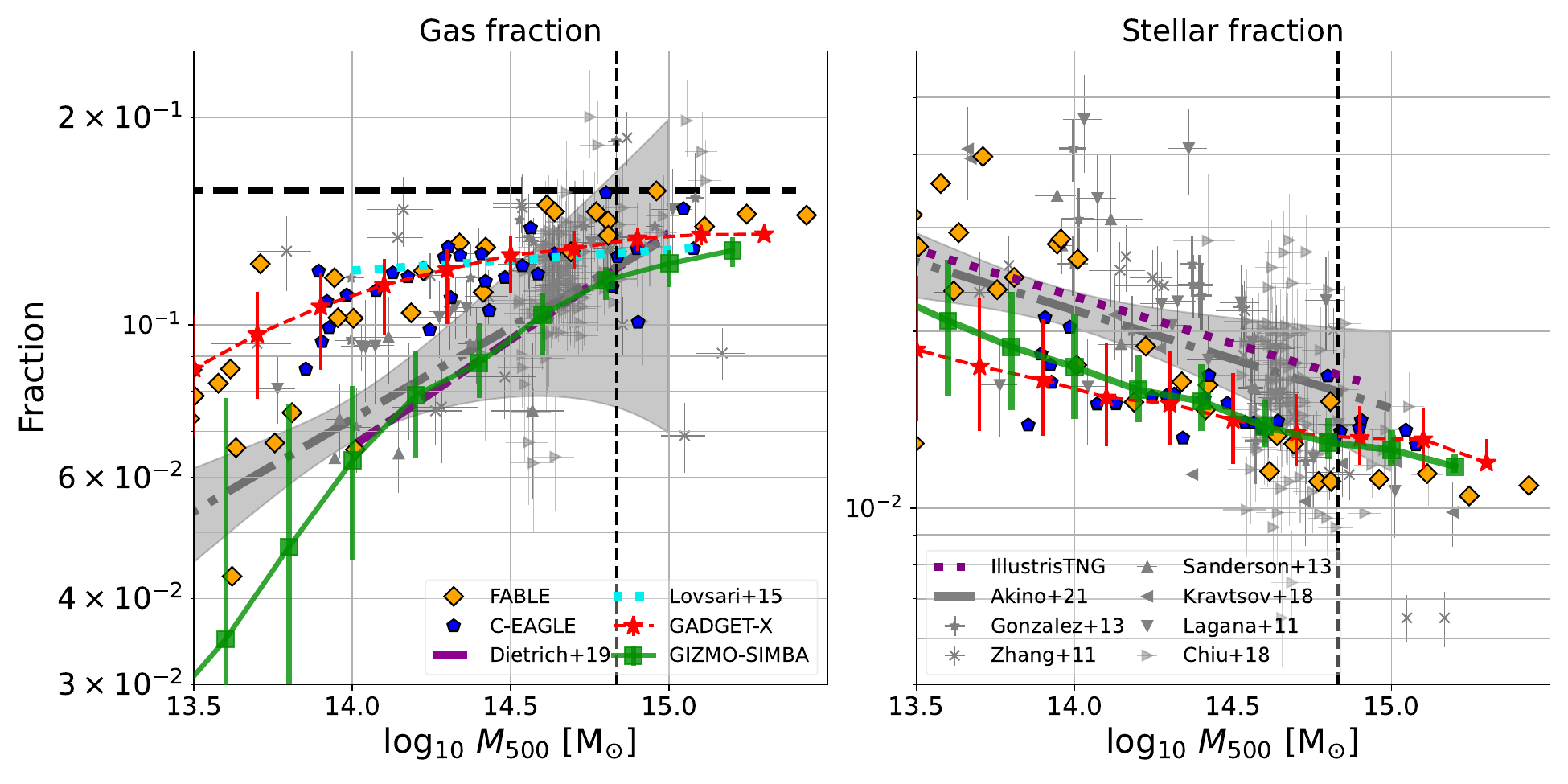}
    \caption{The baryon fractions within $R_{500}$: gas fractions on the left-hand side panel and stellar fractions on the right-hand side panel at $z = 0$. Observational and different simulation results can be crossly viewed from legends in both figures. The symbols and/or line styles for the same reference result are the same in both figures. Therefore, we only show them once in either legend. The statistical results from \theth\ project: \gadgetx\ and \simba, are presented with both symbols and lines. Note that the error bars for the two runs are marking the $16^{th} - 84^{th}$ percentiles. For the statistical/fitting results, we only include the errorbar (light shaded areas in both figures) for the most recent result -- \citet{Akino2021}. The vertical dashed lines indicate the mass completeness for the clusters from \theth\ project \citep[see][for details]{300Cui2022} and the horizontal line in the left-hand-side figure is the cosmological baryon fraction from the Planck cosmology \citep{planck2016}. Note that all the fitting results only cover the region of the observed data points. These two plots show that \gadgetx\ is very similar to the \fable\ and the \ceagle\ simulations in both fractions; there is little difference between \gadgetx\ and \simba\ in the stellar fraction, but the gas fraction from \simba\ shows a much steeper slope. Both fractions from the observational data present a large scatter. }
    \label{fig:1}
\end{figure*}

The first quantity we examine in \simba\ clusters is the abundance of stars and gas. This global quantity provide the overall measure of the cluster's baryonic content that can be directly compared to observational results \citep[see][for a recent review]{Oppenheimer2021}. As have been revealed by both observation and theoretical works \citep[e.g.][]{Behroozi2013b, Yang2013}, more gas is consumed and converted into stars (i.e. star formation efficiency is higher) in less massive halos through the group and cluster regime. Therefore, low-mass halos also tend to have somewhat higher stellar mass fractions. To achieve this general trend, AGN feedback is invoked in all current models~\citep{Somerville2015} in order to solve the cooling flow problem~\citep[e.g.][]{Fabian1994} in which too many stars are formed, especially in the BCG, due to gas cooling being very efficient in the centres of galaxy clusters \citep[][]{Kravtsov2012}.  In models, AGN feedback from the BH of the central galaxy is invoked to counteract cooling and/or expel gas, thereby quenching the galaxy.  Using this, hydrodynamic simulations of galaxy clusters can roughly reproduce the correct stellar mass fraction as a function of halo mass.

In the case of \simba, we have used the stellar fraction to constrain our baryon parameters (C1).  But since the calibration was only done for a single object, it is still interesting to examine this relation over all the \simba\ cluster regions.  The gas fractions were not used for calibration, so they represent an independent prediction.

In \autoref{fig:1} we present comparisons of both stellar and gas fractions within $R_{500}$ between our simulated clusters at $z=0$, versus recent simulations (\fable, \citealt{Henden2018} and \ceagle, \citealt{Barnes2017b}) and observational data at $z \lesssim 0.1$ \citep{Zhang2011,Lagana2011,Sanderson2013,Gonzalez2013,Kravtsov2018,Chiu2018}. The gas fraction is calculated using all gas particles, but as indicated in \cite{300Li2020}, the cold gas only contributes a very small faction to the total gas mass, hence it is reasonable to compare to the results from observations which mainly use hot gas. Statistical or best-fit results are presented for \citet{Lovisari2015,Eckert2016,Dietrich2019} since these do not provide individual cluster data. Note that other works, such as \cite{Lin2004b,Gonzalez2007,Andersson2011,Budzynski2014} and also the recent works \cite{Lim2020,Chen2022} (estimating gas fractions with SZ signal instead of X-ray) that also examined clusters at $z \sim 0.1$ to investigate similar fractions, are not included in this comparison due to \Romeel{various} reasons, such as no $M_{500}$ or $M_*$ being available, or that the sample is dominated by $z\gg 0$ clusters. However, it has been suggested that there is almost no redshift evolution in both gas and stellar fractions within $z \sim 1$ (see more discussion below), so we include the best fitting results from the recent work of \cite{Akino2021} out to modest redshifts using weak-lensing masses \citep{Umetsu2020} for the 136 XXL clusters in the HSC-SSP survey, shown as the grey shaded band.

As seen by comparing the gas fractions in the left panel of \autoref{fig:1}, among the simulations \gadgetx, \fable\ and \ceagle\ are in very good agreement with each other for both the median and scatters. In contrast, \simba\ shows a steeper slope, steepening further for $M_{500} \la 10^{14} M_\odot$.  \Romeel{A similarly steep trend was found within the group regime in the original {\sc Simba} simulation \citep{Robson2020}.} The difference could owe to the fact that the first three simulations employ a thermally-based AGN feedback scheme, while \simba\ employs a kinetic scheme; we leave a detailed exploration into the origin of such differences for future work.

Turning to the gas fraction observations, both individual clusters and sample fits show a wide range of gas fractions particularly towards lower masses. All the simulation predictions are contained within the observational scatter, but the scatter in any given model is much smaller than in the data. The fitting function from \cite{Lovisari2015} is the flattest among observations, and agrees well with \gadgetx\ results. Meanwhile, \cite{Dietrich2019} and \cite{Akino2021} are in  better agreement with \simba\ at $M_{500} > 10^{14} \Msun$; measurements to even lower masses are as yet highly uncertain. The downturn for $M_{500} < 10^{14} \Mstar$ in \simba\ owes to the high jet velocity which blows gas particles well outside of these low-mass halos' virial radius~\citep{Sorini2021}. 
While these observations (and to a lesser extent the simulations) assume slightly different cosmological parameters, most are broadly consistent with a {\it Planck} cosmology \citep{planck2016}. Given the complexity in deriving these observed gas masses and $M_{500}$, it is not obvious how to make a correction for cosmology, so we take the data as-is and simply note that the predicted and observed ranges are likely to be much larger than differences due to cosmology.

The right-hand panel of \autoref{fig:1} shows the stellar mass fractions versus $M_{500}$.  Here, \gadgetx\ and \simba\ show very similar results, and IllustrisTNG \citep{Pillepich2018} shows a similar slope but a slightly higher amplitude that is in better agreement with the observations from \cite{Akino2021}.
In contrast, \fable\ shows higher stellar fractions at group scales ($M_{500}<10^{14}\Msun$), and thus a steeper slope that matches better with the individual observations shown \citep[also the results in][]{Andreon2010}.  We note that the low resolution of \theth\ will tend to suppress stellar fractions since they do not resolve as far down the mass function as IllustrisTNG; if the shape of the galaxy stellar mass function is not a strong function of $M_{500}$, this can explain the constant $\sim 20\%$ offset between these models.  Overall, we note that where the observations are most robust at $M_{500}\ga 10^{14.5}\Msun$, all models are within the range of the observations. The scatter in the simulations is lower than in observations, which may reflect observational uncertainties in addition to intrinsic scatter.  Finally, we note that \cite{Anbajagane2020} compared several cosmological simulations -- BAHAMAS, Magneticum and TNG -- and found that TNG has the lowest total stellar mass within $R_{200}$. Thus, we suspect that the stellar mass fractions from BAHAMAS and Magneticum would also lie above the TNG line in the right panel of \autoref{fig:1}.

In conclusion, for massive ($\sim 10^{15}M_\odot$) clusters all simulations are generally in agreement with each other for both gas fraction (differences within $\sim$5 per cent) and stellar fractions (difference within $\sim 1$ per cent), and in reasonable agreement with observations. For \simba, the latter was mostly achieved via tuning parameters as described in \S\ref{sec:simba}. The models tend to differ more significantly towards lower mass halos, with \simba\ producing particularly low gas fractions in groups, and lower stellar fractions compared to observations of individual $10^{14}M_\odot$ systems. The origin of the differences between \gadgetx\ and \simba\ are being investigated by studying their density profiles in Li et al. (in preparation). Observationally, deeper and more precise estimation of the gas fraction from next-generation surveys such as NIKA2 \citep{Adam2018}, CMB-S4 (Abazajian et al. 2016) using the Sunyaev-Zeldovich effect\footnote{Interested readers are referred to \cite{Yang2022} on how the next-generation SZ observations can be used to distinguish different baryon models.} and ATHENA \citep{Nandra2013} and Lynx \citep{Lynx2018} in the X-rays, will be required to test the input galaxy formation physics. In the meantime, we can look into other properties to distinguish between these simulations. 

\subsubsection{The evolution of the baryon fractions at the same halo mass}
\begin{figure*}
	\includegraphics[width=\textwidth]{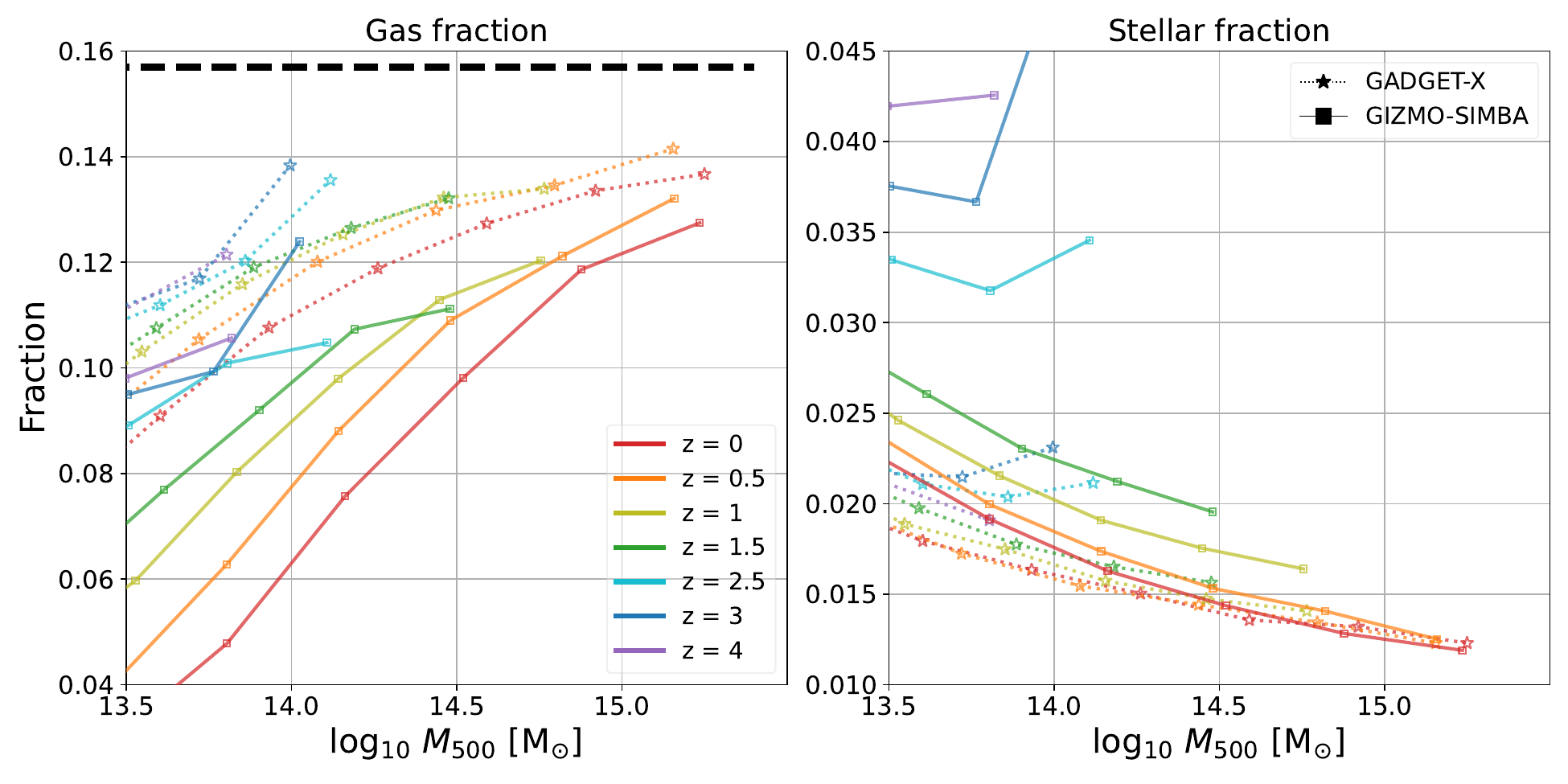}
    \caption{The gas fraction (left-hand-side panel) and stellar fraction (right-hand-side panel) binned by $M_{500}$ at different redshifts. To clearly view the fraction changes along redshift, we use linear scale for the y-axes. The star symbol with dotted lines are for \gadgetx\ and square with solid lines are for \simba. Different colourful lines are used to highlight the redshift evolution, see the legend on the left-hand-side panel for details. We don't include any observation data (especially at high redshift) in this busy plot because there lacks statistical results and weak or no redshift evolution is claimed. It seems that \simba\ shows much clear redshift evolution for both gas (much clearer at lower halo mass end) and stellar fractions compared to \gadgetx, of which shows a weak redshift evolution in their stellar mass fractions.}
    \label{fig:2}
\end{figure*}

It has been suggested that cluster baryon fractions (both gas and stellar) depend weakly on redshift, but strongly on cluster mass (see \citealt{Chiu2018} from observation side, or \citealt{Planelles2013, Truong2018} from simulations). \Romeel{At first glance this seems paradoxical within} hierarchical structure formation models: how can the baryonic scaling relations vary steeply with mass and still remain roughly constant in time as structures grow? In $\Lambda$CDM, big halos are mostly formed later by merging with smaller halos. If we ignore the baryon processes and halo accretion at low redshift, given that the smaller (group-sized) halos have low gas fractions and high stellar fractions, the later-formed massive (cluster) halos should have even lower gas fractions and higher stellar fractions, opposite to what is observed. In this and the following section, we will study the redshift evolution of the galaxy clusters in two ways, by binning at the same halo mass at all redshifts and by tracking individual halos.  The former, \Romeel{discussed in this section,} highlights how baryon fractions change for a sample selected at a given mass, while the latter\Romeel{, discussed in the next section} explicitly shows the \Romeel{true} evolution of baryon fractions as halos grow hierarchically.

For the gas fraction evolution shown in the left panel of \autoref{fig:2}, at all redshifts the clusters show increasing gas fractions with mass, approaching but not reaching the full expected baryonic budget (horizontal dashed line).  There is modest evolution, with the highest gas fractions at high redshifts.  Higher gas fractions are expected at early epochs when cooling is rapid and feedback processes are dominated by star formation whose energetics are typically not sufficient to unbind gas from protoclusters.

Comparing between \gadgetx\ and \simba, the latter has significantly more evolution at lower masses, and overall shows lower gas fractions. 
%This could be caused by its high star formation rate (see the very high stellar mass fraction in the right panel). 
In Li et al. (2022, in prep.) we identify that the gas fraction difference between \gadgetx\ and \simba\ owes to \gadgetx\ tending to have a much higher gas density in the halo centre than \simba.  This points to galactic feedback processes being the primary driver of the model differences.  This is corroborated by the most significant drop occurring between $z=2.5\to 1.5$, which is when {\it Simba}'s jet mode AGN feedback which becomes important prominent in this mass range~\citep{Robson2022}.  The gas fraction continues to drop more rapidly in \simba\ that in \gadgetx\ down to $z=0$ in $M_{500}\la 10^{14.5}M_\odot$ halos.

Although for clarity we don't show higher-redshift observation data in \autoref{fig:2}, \simba\ has the closest gas fractions compared to \cite{Chiu2016a}, around 10 per cent at $z\approx0.9$ and $M_{500} \approx 6\times 10^{14} \Msun$, while the gas fraction from \cite{Chiu2018} (around 12 per cent at $z\approx0.6$ and $M_{500} \approx 4.8\times 10^{14} \Msun$) lies between \gadgetx\ and \simba. 
However, \cite{Chiu2016a} suggests no statistically significant redshift trend at fixed mass \citep[see also][]{Chiu2018,Bulbul2019} when accounting for a 15 per cent systematic mass uncertainty. 
\simba\ predicts some evolution over this redshift range, but at a $\la 1\%$ level which is much smaller than the uncertainties in \cite{Chiu2018}). This is also suggested by \cite{Henden2020} and the redshift evolution is even stronger at poor group masses of $0.5-1\times 10^{14} \Msun$. 

The stellar fraction--halo mass relation at various redshifts is shown in the right figure of \autoref{fig:2}.  Both simulations predict a dropping stellar fraction with mass for $M_{500}\la 10^{14.5}M_\odot$, and by $z=0$ they are fairly similar.  However, the evolution is quite different, with again \simba\ showing much more evolution than \gadgetx.  Neither simulation evolves much at the massive end; we reiterate that \simba\ was calibrated to match $z=0$ observations in this mass range.

The rapid drop in stellar fraction indicates that galaxies in these systems tend to form their stars very early on, so that over time the halo mass grows but the stellar content does not keep up.  This can happen because star formation is quenched early on, and also if the hierarchical growth is predominantly adding smaller systems that have lower stellar fractions (well below group scales). 

For massive cluster halos, the stellar mass should mainly come from the accreting small halos, as the central galaxy is typically quenched by $z\sim3$ (see the following subsection and \S\ref{subsec:bcg_ft} for details). This results in mild but still visible redshift evolution -- about 0.5 per cent from $z=1.5\to 0$.

In smaller clusters and groups, jet feedback in \simba\ is once again implicated by the fact that the most significant change in stellar fraction happens between $z=2.5$ and $z=1.5$.  We have argued that jet feedback is responsible for dropping the gas fraction, which then reduces the fuel for star formation and thus causes galaxy quenching that results a decrease in the stellar fraction.  From $z=1\to 0$ the evolution is relatively modest.  However, it is still significantly more than in \gadgetx, which predicts essentially no evolution in stellar mass fractions since $z\sim 1.5$.  This suggests that the stellar fractions within protocluster environments at high redshifts provides a significant discriminant between models.

Although we don't include observations of higher-$z$ stellar fraction determinations on this plot, we comment on some \Romeel{other} results below in relation to our predictions.
\Romeel{On the simulation side}, the mild evolution is in agreement with \cite{Henden2020} \Romeel{who} also report a marginally significant change for $M_{500} \gtrsim 3 \times 10^{14} \Msun$ up to $z \sim 1.2$. \Romeel{In contrast, at lower masses}, they \Romeel{predict the opposite evolution with redshift versus} \simba.  \Romeel{Among observations,}  although the stellar fraction of 0.023 at $8 \times 10^{13}\Msun$ with the median redshift around 0.5 from \cite{Chiu2016b} is in good agreement (see also \citealt{Chen2021} for a similar result at lower redshift and $M_{200}$) with \simba\, the stellar fractions from \cite{Chiu2018} -- 0.0083 at $M_{500} \approx 4.8 \times 10^{14}\Msun$ and $z\approx 0.6$, and from \cite{Chiu2016a} -- 0.011 at $M_{500} \approx 6 \times 10^{14}\Msun$ and $z\approx 0.9$, are much lower than both \gadgetx\ and \simba. Controversially, \cite{Decker2021} report higher $f_* \approx 0.025$ for 12 clusters at $z=0.95-1.43$, which seems in good agreement with \simba. Besides these results, most observations claim there is no clear redshift evolution of the cluster stellar fraction, \citep[see][for example]{Lin2012,Lin2017,Chiu2018}. This likely owes to small sample sizes, the difficulty in measuring stellar fractions in distant systems, and the larger uncertainty in mass estimation. Therefore, such a very small fraction change predicted in simulations can be difficult to detect. Note that using $M_{500}$ means we don't \Romeel{account for} halo pseudo-evolution, \Romeel{but this is the case for both} observations and simulations so this should not have an effect on our \Romeel{comparisons}. Another point to keep in mind is that the high redshift halos in zoom simulations are necessarily the progenitors of the $z=0$ halos, unlike in observations. Therefore, the redshift evolution can appear more significant than comparing to random samples from a cosmological volume.

In conclusion, we find that there is a  decrease in both gas and stellar fractions with time at lower halo mass ($M_{500} \lesssim 10^{14.5}\hMsun$) in both \gadgetx\ and \simba.  However, the rate of decrease is much stronger in \simba, which shows much higher stellar fractions at high-$z$ and much lower gas fractions at low-$z$. This differences can be expected from \simba's kinetic AGN feedback which imparts much stronger feedback than \gadgetx's thermal AGN feedback. For high mass halos, the AGN feedback becomes relatively unimportant, and both models predict similar gas and stellar fractions at all redshifts \citep[see][for a recent review on the AGN feedback on galaxy groups]{Eckert2021}.  Thus the greatest discrimination between models occurs at high redshifts for poor clusters and groups, which motivates future X-ray and SZ surveys that can probe this regime.

\subsubsection{The evolution of the cluster baryon fractions by tracking}
\begin{figure*}
	\includegraphics[width=\textwidth]{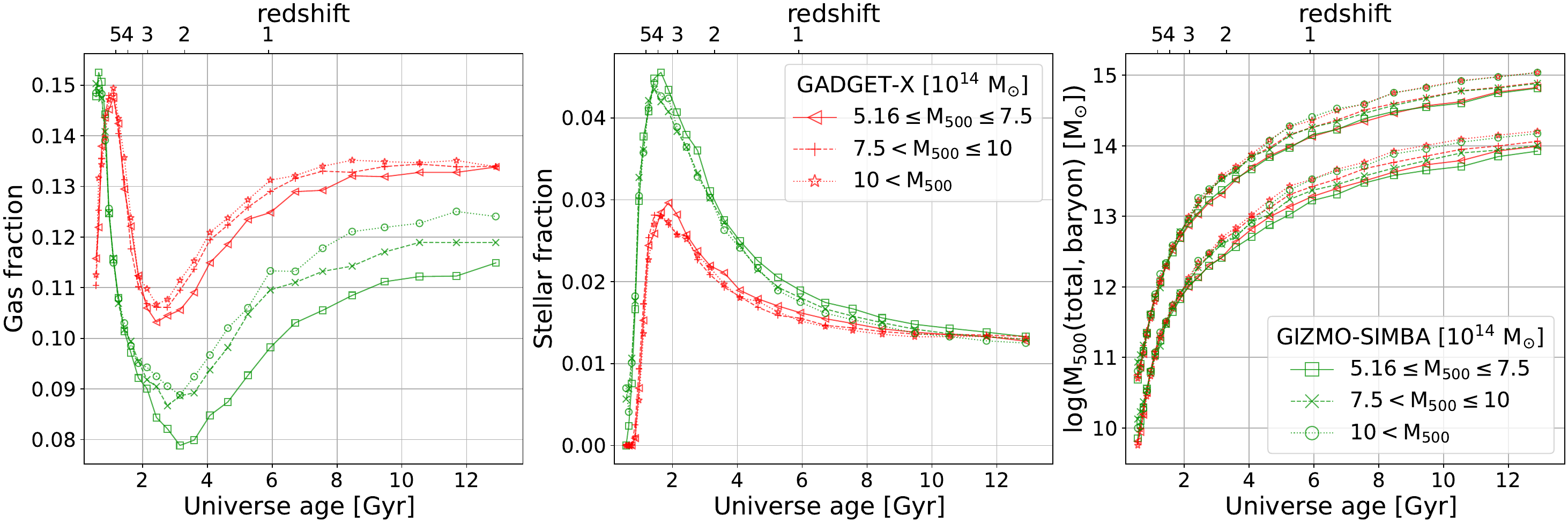}
    \caption{The redshift evolution of gas fraction (left-hand-side panel), stellar fraction (middle panel) and total/baryon masses (upper and lower family of lines in the right-hand-side panel, respectively) within $R_{500}$ for three halo mass bins at $z=0$. The halos at $z=0$ are separated into the same three halo mass bins as indicated in the legends. Only the median values of the halos and their progenitors in the same mass bins are shown. Errorbars are not shown because of the very dense data points and the curves are very close to each other. Through this tracking which also goes the highest redshift in the simulation snapshots, we can view the detailed differences between \gadgetx\ and \simba.}
    \label{fig:3}
\end{figure*}

Simulations provide a view that cannot be \Romeel{seen} in observations, in that they can track the evolution of individual halos over time.  In comparing baryon fractions at a fixed halo mass over time, we saw a significant change at $z\sim 1.5-2.5$ in \simba.  To see how this is reflected in individual halos, we can place clusters into three mass bins at $z=0$, and then track the individual systems back to the earliest epochs, showing how a specific set of clusters evolves.  We track each cluster's main progenitor using the \ahf\ {\sc MERGERTREE} catalogue.  We show their median baryon fraction values as a function of time (and redshift) in \autoref{fig:3}, for our three chosen mass bins as indicated in the legend, for \simba\ (green curves) and \gadgetx\ (blue). 

The gas fraction evolution is shown in the left panel of \autoref{fig:3}.  \gadgetx\ and \simba\ show qualitatively similar evolution, but quantitatively there are differences in the values as well as shifts in the maxima and minima of the evolution. At very high redshift ($z \approx 5-6$), \gadgetx\ and \simba\ reach a gas fraction near the cosmic baryon fraction value, with \simba's occurring slightly earlier. The lower values prior to this likely owe to the halos not being properly resolved at these very early epochs.   From then until $z\sim 2$, gas fractions starts to decline very quickly (more strongly in \simba), until they reach a minimum at $z\sim 2-3$ (slightly later in \simba).  At these redshifts, the feedback models are relatively simple, since there is little AGN feedback and wind recycling is not yet common. Hence differences between \simba\ and \gadgetx\ are caused by the different star formation and stellar feedback models.  The net result is that the gas fraction in \simba\ is only slightly lower (a few per cent) than \gadgetx, persisting until today, indicating that the sensitivity to feedback for halo gas contents is established early on.  After $z\sim 2$, gas fractions start to increase gradually, levelling off after $z\sim 1$ at approximately their present-day value.  

The middle panel shows the stellar fraction evolution. Here, there is sharp early rise as rapid cooling in the dense early universe is able to drive copious gas to the halo centres. After $z\sim 4$, the stellar fractions start to drop, and level off after $z\sim 1$. \simba\ has higher stellar fractions at early times versus \gadgetx; at $z=4$, \simba\ has produce $\sim 50$ per cent more stars than \gadgetx\ in these halos, but by $z=0$ they are the same.  This shows that \simba\ has earlier stellar formation times in groups and clusters compared to \gadgetx. \simba\ grows early galaxies faster than \gadgetx. 

Comparing these first two panels, it is clear that the sharp early drop in gas content accompanies the sharp increase in stellar content.  This implicates rapid gas consumption owing to star formation as the main driver of these early trends.  At later epochs, the lower gas contents in \simba\ limit the available fuel for star formation, resulting in less rapid stellar growth, and the stellar fraction returns to meet \gadgetx\ by $z=0$.

Comparing the three mass bins, until 2 Gyrs there is little difference among the mass bins in either run, in either gas or stellar fractions.  This shows that at early times, galaxies within these regions grow self-similarly along a linear SFR-$M_*$ relation, so there is no significant trend with halo mass.  But as time goes on, the galaxies in more massive halos break away from this as they quench owing to AGN feedback.  These differences are most apparent in \simba, where the low-mass halos have lower gas fractions because AGN feedback can more easily remove baryons. Interestingly, the stellar fractions show much less trends with halo mass. This is because once these halos quench at $z\sim 2-3$, the stellar content doesn't grow much, while the halo mass continues to grow.

The growth of the total and baryon halo mass is illustrated in the right panel of \autoref{fig:3}. As expected, the total halo masses show no differences in evolution between the two runs, as this is driven by hierarchical structure formation. There is about 2 orders of magnitude growth of the total halo mass in the first 2 Gyrs, while the rest of the growth in halo mass (also about 2 orders of magnitude) takes the rest 10 Gyrs.  The total baryon mass shows more differences, owing to \simba\ having lower gas fractions.

The differences in the gas fractions (and associated slight differences in the stellar fractions) emerging at $z\sim 2-3$ can be understood from an interplay between halo growth and star formation quenching.  As shown in \cite{Cui2021,Robson2022}, it is at these redshifts when \simba's jet mode AGN feedback turns on in these massive halos, which is what drives galaxy quenching~\citep{Dave2019}. With star formation curtailed and the halo having a stable virial shock~\citep{Keres2005,Dekel2006}, accreted gas can be held up and accumulated in the halo, causing the gas fractions to increase and the stellar fractions to drop. \simba\ and \gadgetx\ have their AGN feedback in operation at a similar time, which is what is required to produce today's quenched galaxy population. Therefore, both models present a qualitatively similar picture, although the quantitative details of their evolutions differ.

In summary, by tracking individual halos within mass bins, we see that the differences in stellar and gas fractions between \gadgetx\ and \simba\ are established at fairly early epochs.  The gas fractions, remain are significantly different at all redshifts with \simba\ having low gas fractions at about $98$ per cent of the cosmic baryon fraction.  \simba\ also shows a stronger dependence with halo mass owing to the ability of its AGN jet feedback model to evacuate gas particularly from smaller halos.  Meanwhile, \simba\ has an earlier stellar growth phase relative to \gadgetx, with $\sim 50\%$ more stars formed than in \gadgetx\ at $z\sim 3-5$. This is caused by \simba\ consuming more gas into stars.  After this, the stellar fractions drop in both models, as AGN feedback quenches galaxy stellar growth while the halo mass continues to grow.  The lower gas fractions in \simba\ curtail stellar growth more than in \gadgetx, so that by $z\la 1$ the stellar fractions in the two models are similar.  The overall differences in the halo baryon masses are small but noticeable by $z=0$.  

While both models produce a similar global stellar fraction by $z=0$, they arrive at this by different histories.  Hence next we examine BCG evolution in more detail to understand how such differences manifest in their stellar properties.

\subsection{BCG Properties}

We now look into the key stellar component in the galaxy clusters: the brightest cluster galaxy. We will study several relations \Romeel{for} BCGs, from their black holes to their halo properties. We further look into the BCG colour--magnitude and age--halo formation time relations. Note that various definitions of BCGs are used in this section at times, to better compare with observational results. 

\subsubsection{C2: The $M_{BCG, *}$ - $M_{halo}$ relation}
\begin{figure*}
	\includegraphics[width=\textwidth]{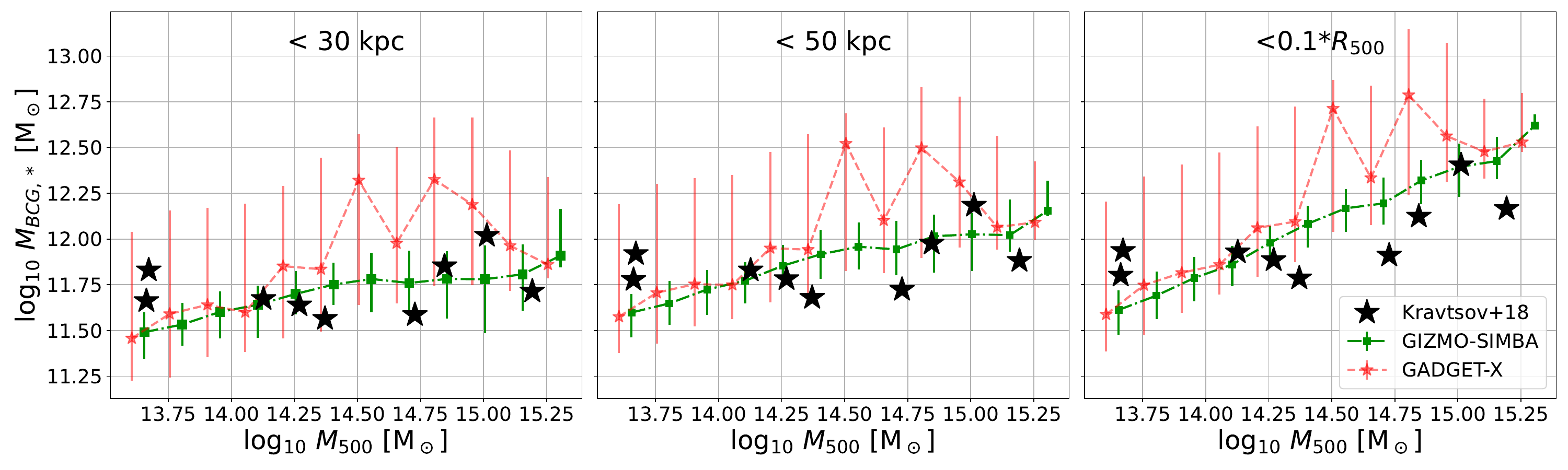}
    \caption{The BCG mass -$M_{500}$ relation from the \gadgetx\ and \simba\ runs of \theth. From left to right panels, we show the BCG mass estimated within different 3D apertures: 30 kpc, 50 kpc and 0.1$*R_{500}$. The observation results from \citet{Kravtsov2018} are shown by black filled stars. The red, blue and green lines with error bars are the median values with $16^{th} - 84^{th}$ percentiles from \gadgetx\ and \simba, respectively. Generally, \gadgetx\ and \simba\ are inline with the \citet{Kravtsov2018}.}
    \label{fig:4}
\end{figure*}

As we have discussed in \autoref{sec:simba}, the BCG -- halo mass relation is one of the three constraints for the \simba\ parameter calibration. One well-known issue for the BCGs in galaxy clusters is that they intermingle with the intra-cluster light (ICL) in their outskirts. It is not trivial to separate a BCG's stars from its surrounding ICL in hydrodynamic simulations \citep[see][for examples]{Murante2007,Dolag2010,Canas2020}. A new advanced method has been proposed in Ca\~nas et al. 2022 (in prep.), which is robust and independent of simulation resolution. However, the BCG identified with these more physical motivated methods is not directly comparable with the BCG as identified in observations \citep[see][for more discussions]{Cui2014b,Rudick2011}. 
Therefore, it is not easy to make apples-to-apples comparisons. Here, we adopt a simple definition of the enclosed mass within different 3-D apertures, to be able to compare with the observation results from \cite{Kravtsov2018}. Note here that the observed masses are within  projected 2D apertures, so the simulations are expected to have a slightly lower BCG mass compared to the observed one.

% Most of results are based on halo mass of $M_{200}$, such as \cite{Zhang2016}, or the BCG mass is not calculated within an aperture, such as \cite{Lidman2012}, while the result from \cite{DeMaio2018} is based on clusters at $z\gtrsim0.3$. It has been suggested that the BCG mass growth after $z=0.3$ is at around 20 per cent \citep[for example][]{DeLucia2007,Guo2011}, which, however, seems not support by observation \citep[see][for details]{lin_stellar_2013,Zhang2016,Cerulo2019,DeMaio2020}. This suggests the notion that BCGs had accreted most of their mass by $z \approx 0.3$. Note that \cite{Zhang2016} claimed that this tension can be released when taking ICL evolution into account in SAMs\citep{Contini2014}. Later hydrodynamic simulations \citep[for example][]{Ragone-Figueroa2018,Henden2020} found a similar evolution as suggested by observation. For simplicity, we don't include these results in this comparison. However, as shown in Fig.5 of \cite{Henden2020}, these observation results are very similar to \cite{Kravtsov2018}. 
%{\bf WEIGUANG: I'm confused why are you discussing all these observations/models if we aren't going to include them (also I'm not sure which are model results and which are data).  I'm really not sure what I'm supposed to learn, especially with all the contradictory statements.} This is mainly about the BCG evolution, which i was thinking as interesting to be included for dicussion.

In \autoref{fig:4}, we show the BCG stellar mass--$M_{500}$ relations from all the uncontaminated halos in our 324 regions at $z=0$. Here, the BCG stellar masses are estimated at three different 3D apertures: 30 kpc (left), 50 kpc (middle) and $0.1\times R_{500}$ (right). \simba\ results are in green, \gadgetx\ in blue, and observations from \citet{Kravtsov2018} are shown as black stars.  Errorbars enclose 16-84\% of the simulated values.

Overall, there is broad agreement with the \citet{Kravtsov2018} data for both models, with a mild increase in BCG mass as a function of $M_{500}$.  The general trends are very similar regardless of the chosen aperture.  However, in detail there are some differences.  Generally, \simba\ produces somewhat lower BCG masses, particularly at $M_{500} \lesssim 10^{14} M_\odot$.
Also, \gadgetx's BCG masses show a clearly larger dispersion at a fixed halo mass bin.
At all apertures, the BCG masses from \simba\ are in somewhat better agreement with the \cite{Kravtsov2018} data. 
\Romeel{To quantify aperture effects for 3D versus 2D,} we have checked the projection effect on the BCG masses within different apertures and find that the BCG mass increases are \Romeel{mostly less than} 20 per cent. This increase will not change the agreement with the BCG masses from \cite{Kravtsov2018}.
It seems that both models predicts a slightly steeper slope for the $M_{BCG, 0.1\times R_{500}} - M_{500}$ relation than \cite{Kravtsov2018}, which could be caused by different amount of ICL contribution which will be studied in detail in Ca\~{n}as et al. 2022 (in prep.).

As shown in \cite{Henden2020}, the clusters from \fable\ simulation have a systematically higher BCG stellar mass compared to observations (by $\sim$0.2-0.3 dex, which is still a significant improvement from \citealt{Puchwein2010}). The IllustrisTNG and C-EAGLE clusters (orange dashed line and orange circles in Fig. 5 of \citealt{Henden2020}, see also \citealt{Pillepich2018}) also possess significantly more massive BCGs than observed despite their more realistic galaxy sizes. Furthermore, \cite{Anbajagane2020} indicated that the BCG mass (within 100 kpc) from Magneticum is even higher than IllustrisTNG, while BAHAMAS seems to have a slightly lower stellar mass compared IllustrisTNG. \cite{Ragone-Figueroa2018} show that their BCG masses are in very good agreement with observations, which could be the improvement on the BH's cold and hot modes gas accretion compared to their previous result \citep{Ragone-Figueroa2013}. \Romeel{In} comparison to these works, the BCG \Romeel{masses} in \simba\ \Romeel{seem somewhat more} realistic.

\subsubsection{BH--BCG--halo relations}
\begin{figure*}
	\includegraphics[width=\textwidth]{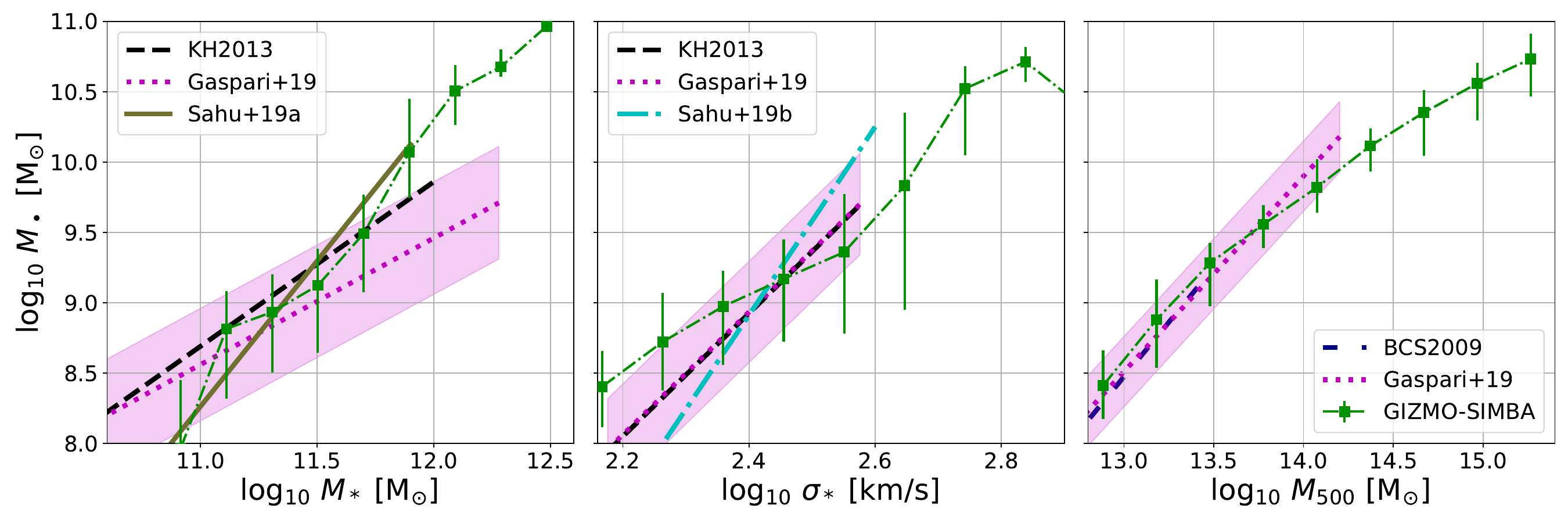}
    \caption{The BH -- BCG -- halo relation comparisons between the \simba\ run and observations. From left to right panels, we show the BH mass as a function of BCG stellar mass $M_*$, BCG stellar velocity dispersion $\sigma_*$ and halo mass $M_{500}$, respectively. We refer to the legends in each panel for the detailed observation result with the \simba\ run is also shown in green dotted-dashed lines. Again, the errorbars show $16^{th} - 84^{th}$ percentiles. Shaded regions are the errorbars coming from \citet{Gaspari2019}. Note that \citet{Bandara2009} -- BCS thereafter, used halo mass $M_{200}$ in their fitting result, we simply convert the $M_{200}$ to $M_{500}$ by assuming a fixed concentration of 3. Also note that all the fitting results only roughly cover the region of the observed data points.}
    \label{fig:5}
\end{figure*}

All BCGs are believed to contain large supermassive black holes (BH).  As shown by previous studies  \citep[e.g.][]{Martizzi2012,Martizzi2014,Frigo2019}, AGN feedback is a key to have simulated galaxy clusters reproduce observed ones. Since AGN feedback is tied to the BH accretion rate which also sets the final BH mass, comparisons of the BH--galaxy--halo relations will help us to test and constrain the AGN feedback in models. 

In \autoref{fig:5} we compare \simba's BH mass $M_\bullet$ vs. stellar mass $M_*$ (left), stellar velocity dispersion $\sigma_*$ (middle), and $M_{500}$ (left) versus observations.  Note here $M_*$ and $\sigma_*$ are obtained from the \caesar\ catalogue instead of in a fixed aperture as in the previous section. In addition, only central galaxies are considered in \autoref{fig:5}; i.e. no satellite galaxies are included. \gadgetx\ is excluded from this comparison. For the observation results, we show data for early-type massive galaxies from \cite{Gaspari2019} (with BH properties from \citealt{vandenBosch2016}), \cite{Sahu2019a}, \cite{Sahu2019b}, and \cite{Bandara2009}, which are similar to or supersede various other data  \citep{McConnell2013,Savorgnan2016b,Bogdan2018,Kormendy2013,Savorgnan2016a}.
While we don't specifically select early-type galaxies, we \Romeel{confirmed} that our BCGs all \Romeel{have high bulge-to-total ratios and very low rotational support, which is expected} since they are massive, quenched and appear bulge-dominated.

The $M_\bullet - M_*$ relation is shown on the left panel of \autoref{fig:5}.  \simba\ shows good agreement with \cite{Gaspari2019} (and \citealt{Kormendy2013}, not shown) at $M_* \approx 10^{11.2} - 10^{11.7} \Msun$. However, the BH masses for galaxies for $M_* \ga 10^{11.7} \Msun$ galaxies are above these observations. In contrast, \cite{Sahu2019a} found a steeper result that is in very good agreement with \simba.  It is beyond the scope of this work to delve into the differences between these observations results, so we simply note that \simba\ predicts that more massive galaxies will follow the same trend from \cite{Sahu2019a}, all the way up to $M_* \approx 10^{12.5} \Msun$.

Looking at the middle panel of \autoref{fig:5}, there is more deviation between the \simba\ run and the observation results for the $M_\bullet - \sigma_*$ relation. At $\log_{10} \sigma_* \lesssim 2.6$, \simba\ has a flatter slope compared to the observed results. Although earlier observations indicate a shallower slope \citep[e.g.][]{Sabra2015}, the most recent observations prefer a steeper relation; for example \cite{Dullo2021} predicts a even higher slope than \cite{Sahu2019b} for early-type galaxies. Note that \cite{Thomas2019} showed that the $M_\bullet - \sigma$ relation from the original {\sc Simba} simulation is in very good agreement with \cite{Kormendy2013}. This suggests that the re-tuned AGN feedback and lower resolution has more effects on internal dynamics than on the galaxy stellar mass -- it seems to reduce the galaxy velocity dispersion at lower stellar or halo masses. Meanwhile at  $\log_{10} \sigma_* \gtrsim 2.6$, the $M_\bullet - \sigma_*$ relation from \simba\ is basically following the extensions of \cite{Kormendy2013,Gaspari2019}.

There has been some controversy regarding the connection between BH mass and host halo mass, with some claiming a tight relationship \citep{Booth2010,Bogdan2015,Bogdan2018,Bassini2019,Marasco2021}, but others claiming it is incidental \citep[e.g.][]{Kormendy2011} and are only related to classical bulges.   For instance, \cite{Zhang2021} suggests that the BH and AGN feedback is not relevant for the formation history of massive star forming disks which show very high gas-to-star conversion rates, suggesting small BHs in their large halos.  
Nevertheless, we argue here that the $M_\bullet - M_{\rm halo}$ relation exists for these BCGs in galaxy clusters which are mostly ellipticals. Hence examining the BH mass--halo mass connection in \simba\ is interesting.

The result for $M_\bullet-M_{500}$ from \simba\ is shown in the right panel of \autoref{fig:5}.  We compare to \cite{Bandara2009} (BCS2009) and \cite{Gaspari2019}.  \simba\ is in very good agreement with the observed ones at group scales. However, there is a hint that at galaxy cluster scales, \simba\ predicts that the slope that becomes slightly shallower; observations do not extend into this regime. This is understandable as the BCGs in more massive clusters get quenched earlier (see more discussions in \autoref{subsec:bcg_ft}), so like with the stellar mass, the halo continues to grow without commensurate growth in the black hole.  \Romeel{Interestingly, the relation is quite tight with $M_{500}$, even though neither BH accretion nor feedback in \simba\ is directly governed by any halo property.}

% \cite{Huang2020} found that Massive galaxies with more extended stellar mass distributions tend to live in more massive dark matter haloes. The two-parameter $M_{\star }^{max}-M_{\star }^{10}$ description provides a more accurate picture of the galaxy-halo connection at the high-mass end than the simple stellar-halo mass relation (SHMR) and opens a new window to connect the assembly history of haloes with those of central galaxies. The model also predicts that the ex situ component dominates the mass profiles of galaxies at $r < 10$ kpc for $log M_* \ge 11.7$.

\subsubsection{BCG colours} \label{subsec:colour}
\begin{figure}
	\includegraphics[width=0.5\textwidth]{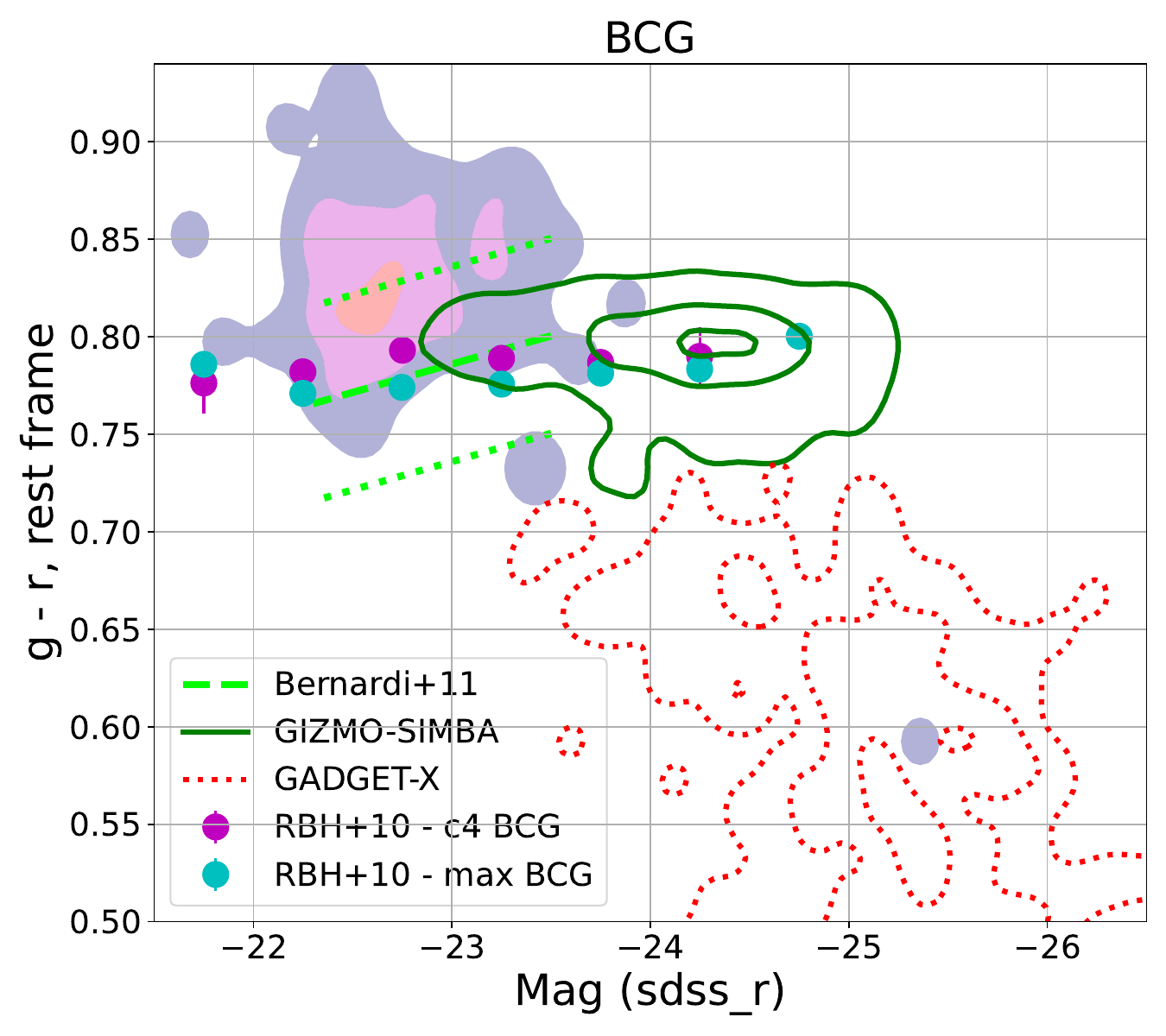}
    \caption{The g - r colour - magnitude diagram for the BCGs in the mass-complete cluster sample compared to the observation results in cluster mass range. The filled colourful contours are the results from SDSS \citep{Yang2018}. All the other simulation and observation data can be deciphered from the legend. The contours for the SDSS, simulation and SAM results are at the same percentiles: $16^{th}, 50^{th}$ and $84^{th}$.}
    \label{fig:6}
\end{figure}

BCG galaxies tend to be red in colour, indicating predominantly older stellar populations.  The colour thus provides a constraint on the formation history of the BCG's stars.  We determine the photometry of the galaxies in \simba\ and \gadgetx\ using the {\sc Pyloser} code implemented in \caesar, which uses the age and metallicity of each star within a galaxy to generate a spectrum based on a simple stellar population model from {\sc FSPS}\footnote{https://github.com/cconroy20/fsps}\citep{Conroy2009,Conroy2010}. We adopt a \citet{Chabrier2003} initial mass function for the stars, matching the one used in the simulation. {\sc Pyloser} also computes the dust extinction to each star based on the line-of-sight dust column, but for BCGs the dust model has a negligible effect since they have little cold ISM gas or dust.

In \autoref{fig:6}, we compare the distributions of the BCG's $g-r$ colour as a function of its magnitude in SDSS $r$-band. The observational results are taken from \citet[best-fit shown as lime dashed line with dotted lines for the rms scatter]{Bernardi2011}; from \cite{Roche2010} for the results of both the C4-BCG \citep[magenta points from][]{Bernardi2007} and the max-BCGs \citep[cyan points from][]{Koester2007} \Romeel{samples}, and from \cite[coloured contours]{Yang2018} with BCGs from their group catalogue. Note that the $g-r$ colour is corrected in rest frame for all the observation data based on \citet{Chilingarian2012}. 

The $g-r$ colour from the observed galaxies typically $\approx 0.8$, slightly higher for \cite{Yang2018}. This is well in line with the BCG colours from \simba, particularly for the \cite{Roche2010} sample which extends to overlapping magnitudes.  In contrast, \gadgetx\ BCGs have a much bluer BCG colour, typically $g-r\approx 0.5-0.7$. This indicates that BCG colours, and hence their star formation histories, are more realistic in \simba\ than in \gadgetx.

That said, for both \simba\ and \gadgetx, the BCG brightnesses extend up farther than the data.  The C4-BCG and max-BCG can reach the magnitude of $r\approx -24.5$ comparable to the brightest galaxies in \simba, but \gadgetx\ shows BCGs extending to brighter than $r\approx -26$.  For \simba, however, the mild differences can probably be attributed to sample selection, in that \theth\ focuses on only the most massive clusters.
Particularly for \gadgetx, the brighter BCGs go hand-in-hand with the bluer colour, since younger stellar populations are more luminous; this is unlikely to be reconciled by the aforementioned effect.

\subsubsection{Halo and BCG formation time} \label{subsec:bcg_ft}

\begin{figure*}
	\includegraphics[width=\textwidth]{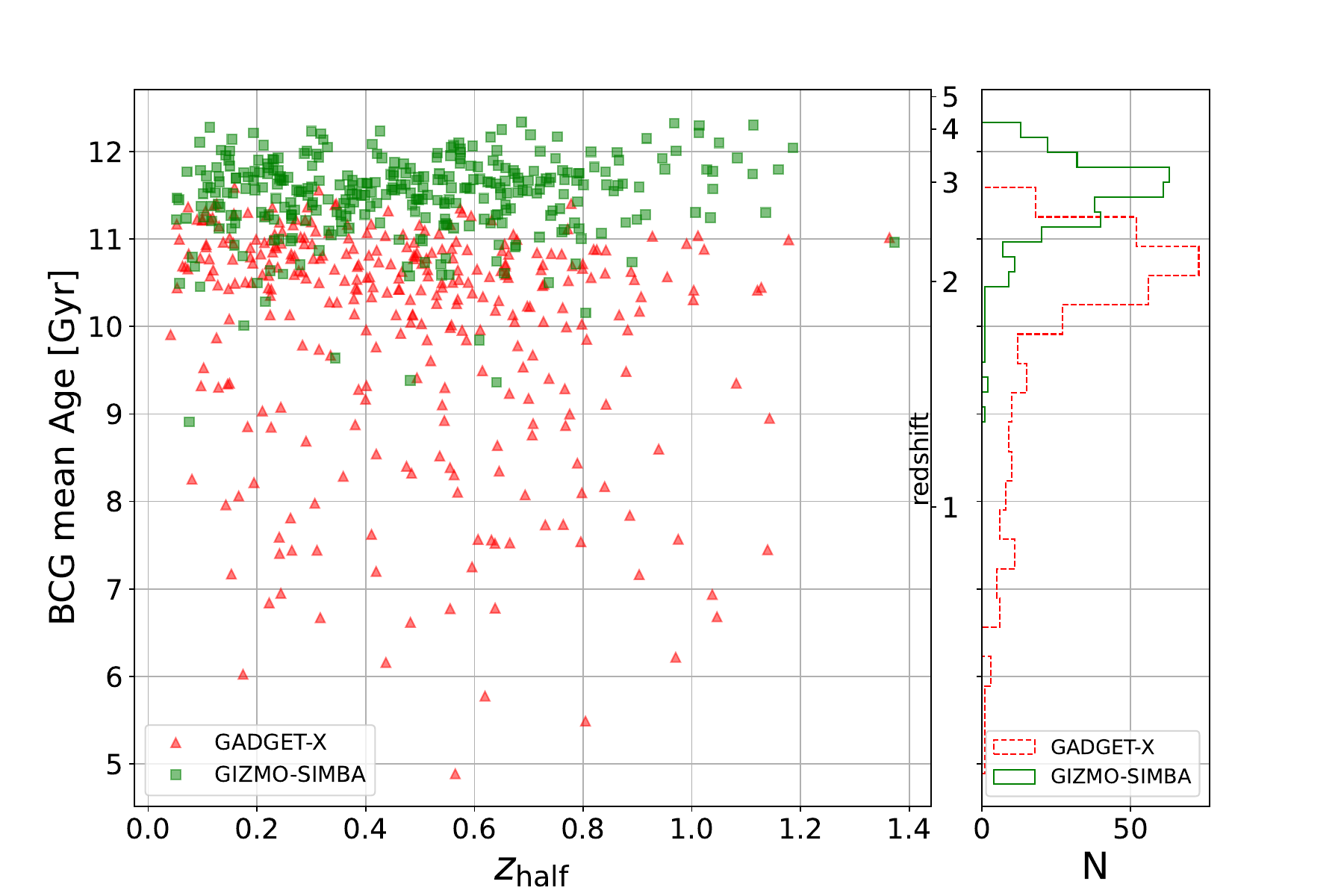}
    \caption{The correlation between the BCG age and halo formation redshift for the mass-complete cluster sample. As indicated in the legend, different symbols are for the BCGs from different versions of simulations and SAMs. We further detail their BCG age distributions by showing the histograms in redshift bins with the same colour to the right side panel.}
    \label{fig:7}
\end{figure*}

The different BCG colours in \simba\ and \gadgetx\ suggest that the BCG \Romeel{stellar} formation times are different in these models.  The $z=0$ stellar populations arise from a combination of the star formation histories in the BCG's progenitors, plus mergers bringing in stellar mass.  Hence in these simulations we can ask two distinct questions: (1) When \Romeel{were} the stars in the BCG at $z=0$ formed? and (2) When \Romeel{were} the progenitor galaxies of the BCG assembled? The first one tightly connects with the detailed star formation history of the BCG, which can be revealed from its mean stellar age; we refer \Romeel{to this} as the star formation time. The second one \Romeel{is} tightly connected to hierarchical structure formation, which \Romeel{is difficult to probe observationally;} we refer \Romeel{to this} as the assembly formation time. We expect this latter one may be more related to halo assembly, which will be quite similar within \simba\ and \gadgetx\ \Romeel{by construction}.

For the first question, the BCG \Romeel{stellar} age has been estimated to be generally very old in observations.  For example, \cite{Edwards2020} found the  luminosity-weighted BCG core age of $\sim$13 Gyr \citep[see also][]{Loubser2009}, which is in agreement with the semi-analytic model prediction from e.g. \cite{DeLucia2007}. However, note that two BCGs (Abell 671 and Abell 602) in the sample of \cite{Edwards2020} have quite young ages of $\sim$ 5 Gyr, which can be caused by more recent in-situ star formation \citep[$\sim$25\% of the growth is in-situ;][]{Ragone-Figueroa2018}. By studying early-quenched galaxies in Magneticum and TNG simulations, \cite{Lustig2022} reported about a factor of 2 older for the average ages of simulated quiescent galaxies than observed ones.

For the second question, one can examine the most massive BCG at high redshift in observations to roughly constrain their assembly time. \cite{Collins2009} found several BCGs at z $\sim 1 - 1.5$ that have stellar masses comparable to the most massive galaxies in the Universe, suggesting a very early assembly time of their BCGs. Similar suggestion is also found in \cite{Andreon2013} which proposed with a even higher redshift $z\sim 2.5$ for the BCG mass build up time. More theoretical studies \Romeel{using} hydro simulations find similar conclusions.  For example, \cite{Ragone-Figueroa2018} found the assembly of the BCG occurs over an extended time-span and half of the BCGs’ stellar mass only falls into place typically by $z\sim 1.5$. Similarly, the BCGs in \fable, which are in good agreement with several observational inferences at $z \lesssim 1$ \citep{Henden2020}, have moderate stellar mass growth, about a factor of 1.5 from $z=1$ to 0.3, and effectively halts at $z \lesssim 0.3$ as suggested by observation. Furthermore, \cite{Rennehan2020} found that the stellar assembly time of a sub-set of brightest cluster galaxies occurs at high redshifts ($z>3$) rather than at low redshifts ($z < 1$). Thus, highly overdense protoclusters assemble their stellar mass into brightest cluster galaxies within $\sim$1 Gyr of evolution, producing massive blue elliptical galaxies at high redshifts ($z>1.5$). \Romeel{Here} we only focus on the first question \Romeel{to determine} if there is any connection between the BCG star formation time and the halo formation time. 

In \autoref{fig:7}, we show the halo formation time versus the BCG star formation time at $z=0$ in \simba\ (green points) and \gadgetx\ (blue).  The star formation time is the mean \Romeel{mass-weighted} age of the stellar population, while 
the halo formation time here is \Romeel{quantified by} the commonly used $z_{\rm half}$, i.e. the redshift by which the halo had accreted half of its final total mass.

Interestingly, there is no obvious correlation between the BCG star formation time and the halo formation redshift, in either model. The halo formation redshifts are fairly recent, generally after $z\sim 1$.  In contrast, the BCG mean stellar age shows the stars are fairly old.  This is a manifestation of stellar population downsizing, where the stars forming in massive halos are old even though the most massive halos assemble late.  This is has sometimes been called "anti-hierarchical" \Romeel{behaviour}, but it is exactly as expected in hierarchical structure formation models where the largest density perturbations collapse first and hence can form the oldest stars, even though the halos assemble late.

In contrast, the BCG age distribution is quite different between \simba\ and \gadgetx.  Clearly, \simba\ has older BCGs, with the formation redshift peaking at $z\sim 3$, and little  star formation after $z\sim 2$. The mean BCG ages from \gadgetx\ is younger,  as young as $\sim 5-6$ Gyrs ago, indicating late star formation which can be caused by either the in-situ star formation shown in \cite{Ragone-Figueroa2018} or rejuvenation at low redshift. This also results in a much larger \Romeel{tail towards young} mean stellar ages in \gadgetx.

These results are consistent with our earlier findings that the stellar assembly within halos is shifted towards earlier epochs in \simba\ vs. \gadgetx.  This owes to both the jet feedback, as well as \simba's X-ray feedback that is responsible for quenching star formation completely \citep{Cui2021}.  In the original {\sc Simba}, the SFR function at $z\sim 2 - 4$ extends to very high SFRs, $>10^3 \Msun yr^{-1}$, which turns out to be critical for reproducing the sub-millimetre galaxy (SMG) population at these epochs \citep{Lovell2021}, which are likely to be the progenitors of BCGs today; other simulations tend to have a more a difficult time matching these constraints.  \Romeel{It seems that} \gadgetx\ may also under-predict the SFR in protoclusters at high redshift, especially the starburst population, compared to observations \citep[see][for more details]{Bassini2020}.  It appears that such rapid early stellar growth within massive halos is crucial both for matching BCG colours today, as well as SMGs at $z\sim 2-4$.

\subsection{Satellite galaxies}

Satellite galaxies provide another important constraint on the stellar properties of galaxy clusters. They can connect with the cluster total mass through cluster richness \citep[for example][]{Anbajagane2020} or velocity dispersion \citep[for example][]{Munari2013,300Anbajagane2022,300Ferragamo2022}, with the cluster's dynamical state through its distribution and mass fraction \citep[for example][]{Cui2017}, and with cluster shapes and orientations. Successfully reproducing the satellite galaxy properties in galaxy cluster is very important for studying environment quenching processes such as ram-pressure stripping and harassment \citep[see][for example]{Lotz2019,Lotz2021}.  Therefore it is important to check how well our simulations reproduce observations of the satellite population in groups and clusters.

\subsubsection{C3: Satellite stellar mass function}
\begin{figure}
	\includegraphics[width=0.5\textwidth]{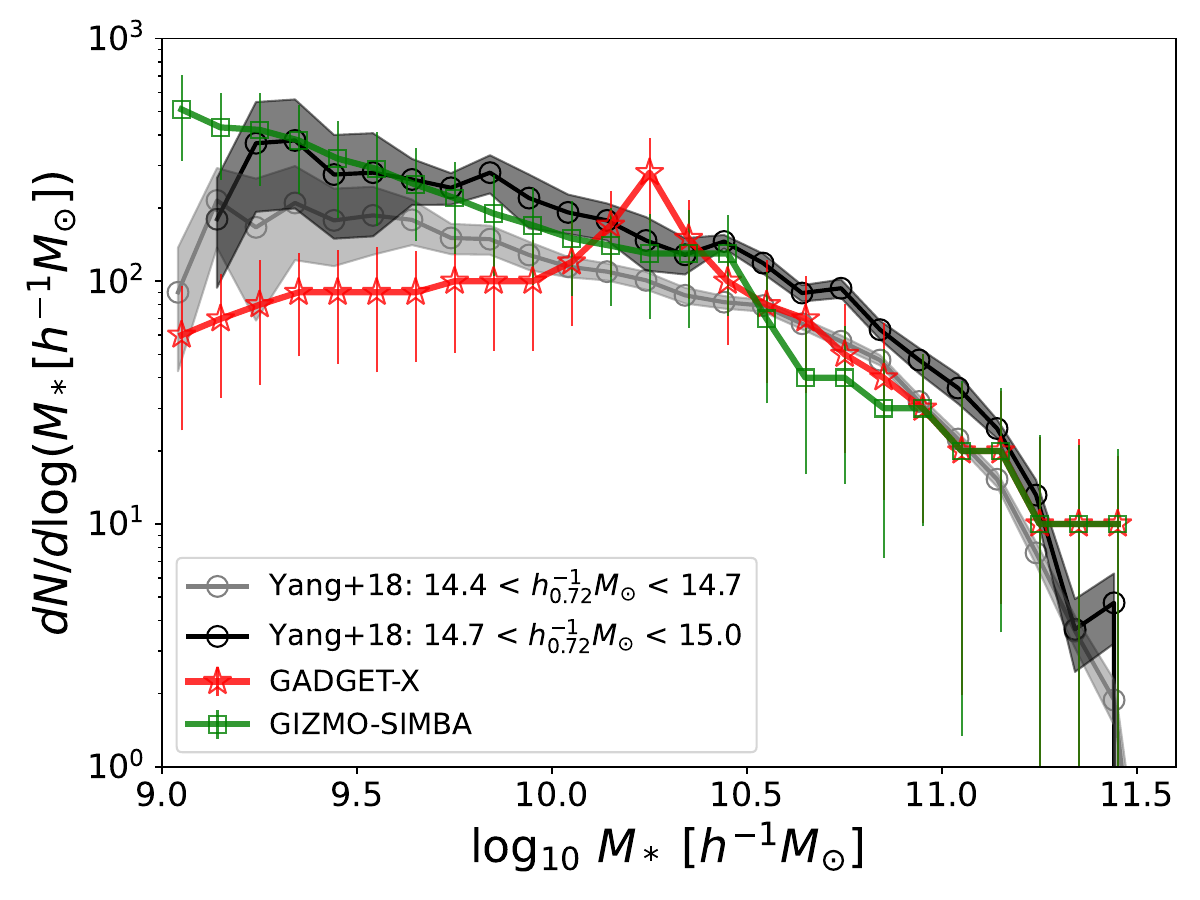}
    \caption{Satellite stellar mass functions. Only the satellite galaxies from the mass-complete clusters are used in this satellite galaxy stellar mass function. We only show the median values in each stellar mass bin of the selected clusters. While the observation results within two particular halo mass bins are shown in black and gray thick lines with errorbars. This plot is different to Fig. 7 in \citet{300Cui2018} by removing the other models results except \gadgetx\ and adding the new \simba\ result with errorbars, which is in very good agreement with the observation results.}
    \label{fig:8}
\end{figure}

We first investigate whether \theth\ simulations produce the correct amount of satellites at different stellar mass bins in \autoref{fig:8}. The satellite stellar mass functions (SSMFs) for \simba\ (green) and \gadgetx\ (blue) are shown, along with observations of the SSMF in clusters from \cite{Yang2018} within two mass bins spanning much of \theth\ sample.
Recall that the satellite stellar mass function is used to constrain the feedback parameters in \simba\ (cf. constraint C3), so the agreement with \simba\ should not be considered a true success of the model, although the tuning was only done on a single cluster.   Detailed comparisons and discussions \Romeel{among} the other \theth\ models can be found in \cite{300Cui2018}. 

The median satellite stellar mass function \Romeel{(SSMF)} from \simba\ show very good agreement with the observation results, which confirms that the tuning down for one region was successful even when considering all 324 regions.   \simba\ and \gadgetx\ agree at $M_*\ga 10^{10} \hMsun$, below which \gadgetx\ falls below the observations.  There are some interesting features, such as \gadgetx\ showing a strong peak at $M_*\ga 10^{10.3} \hMsun$ and \simba\ showing a steep decline at slightly higher masses; we suspect that these arise due to the onset of strong AGN feedback at these masses~\citep{Dave2019}.  This is because the BH is seeded and AGN feedback is in action when the galaxy stellar mass reach $M_* \sim 10^{10.5} \hMsun$, which then halts galaxy growth and piles up galaxies just below this threshold.  A more gradual onset of AGN feedback with halo mass may produce a smoother SSMF; we leave this for a future work. 

Using the Hydrangea simulations, \cite{Bahe2017} reported very good agreement when compared with satellite stellar mass functions from various observations \citep{Yang2009,Vulcani2011,Wang2012}, down to lower masses than we can probe in \theth.  Although \fable\ did not show its satellite stellar mass function, \cite{Henden2020} found that the fraction of the total stellar mass in satellite galaxies is significantly smaller in FABLE clusters ($\sim$40 per cent) than observed. This could be caused by the stars in their satellite galaxies being stripped at an accelerated rate which would also result in a larger ICL mass fraction. It is generally the case that it is easier to strip stars in lower resolution simulations, owing to less deep potential wells, hence this effect may also be relevant for \theth.  \Romeel{Planned higher-resolution runs will be able to quantify this in the future.}

\subsubsection{Satellite galaxy colour distribution}
\begin{figure}
	\includegraphics[width=0.5\textwidth]{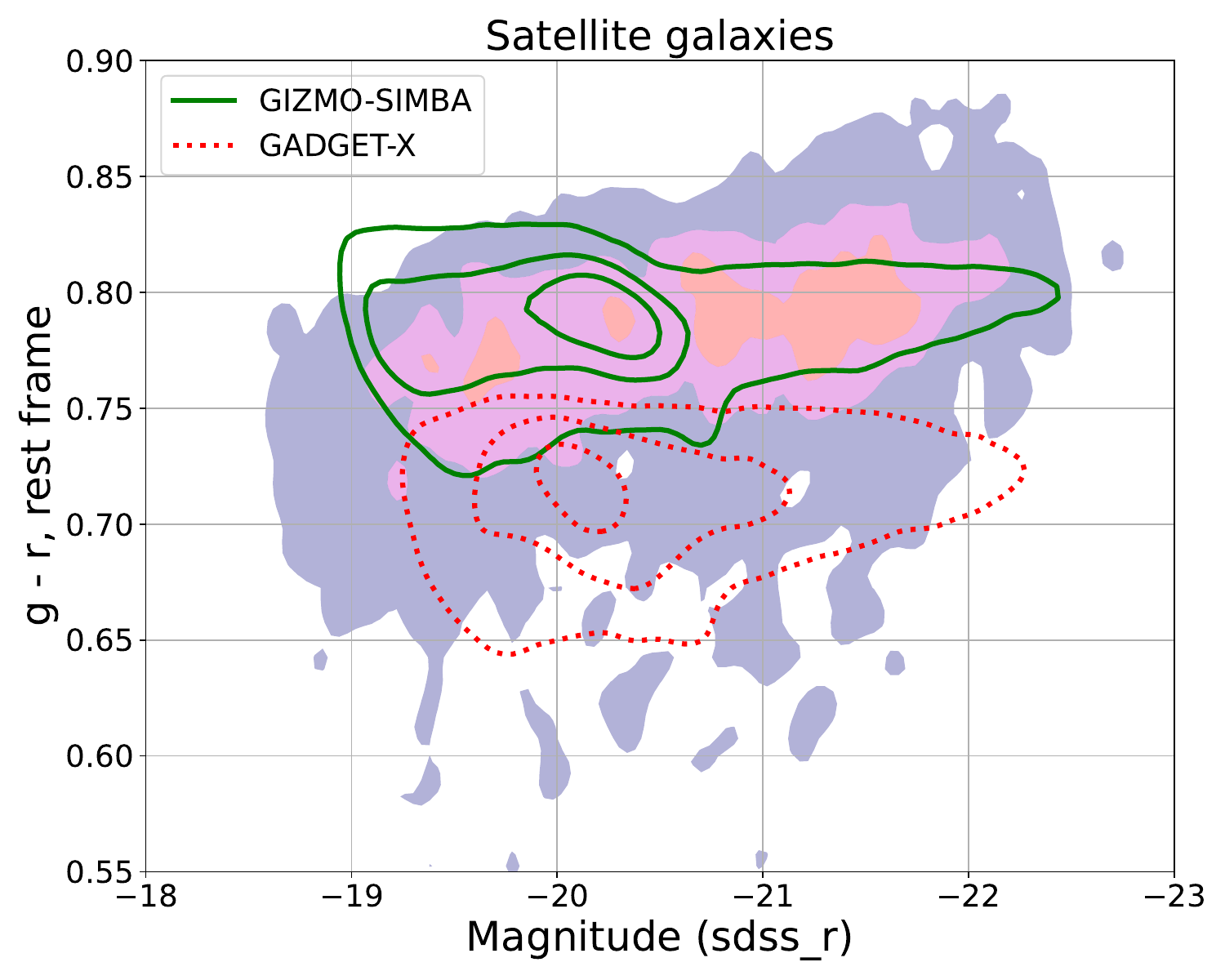}
    \caption{Satellite galaxy colour-magnitude diagram. This plot is similar to \autoref{fig:6}, but for all the satellite galaxies in the mass-complete cluster sample. Contour maps are the observation results from \citet{Yang2018} with clusters $M_{200} \gtrsim 6\times 10^{14} \hMsun$. Again the three contour lines/maps mark the $16^{th}, 50^{th}$ and $84^{th}$ percentiles. Note here that we apply the same stellar mass cut $M_* \geq 10^{10} \Msun$ to both simulation and observed dataset for a consistent comparison.}
    \label{fig:9}
\end{figure}

Similar to the colour-magnitude diagram for the BCGs, the satellite colour-magnitude diagram gives insights into their stellar growth histories.  With \simba\ in particular using the SSMF as a constrain, it is important to see whether it also produces a reasonable distribution of satellite colours as a complementary check on the model.

As shown in \autoref{fig:9}, the $g-r$ colours vs. $r$ magnitudes are shown from \simba\ (green contours) and \gadgetx\ (blue), with the observations from \cite[coloured shading]{Yang2018}.  As with the BCGs, their galaxy colour has been corrected into the rest frame\footnote{Note that this plot is slightly different to the middle panel of Fig. 8 in \cite{300Cui2018} due to a different stellar mass cut ($M_* \geq 10^9 \hMsun$) applied. By comparing to Fig. 8 in \cite{300Cui2018}, \gadgetx\ tends to have a bluer colour, which suggests that less massive satellite galaxies should be dominated by red colour.}. 

Satellite galaxy colours from \simba\ show very good agreement with \cite{Yang2018}, with again typically $g-r\approx 0.8$ just as with the BCGs.  Hence the stellar populations in both the central and satellite galaxies in \simba\ tend to be quite old.  In contrast, \gadgetx\ results, meanwhile, show a somewhat bluer $g-r$ colour, mimicking the trend seen in the BCG colours.  The satellites are somewhat redder than the BCGs in this model.

It is notable however, the \simba's distribution of colours has a small scatter compared to \cite{Yang2018}. In particularly, \simba\ lacks any bluer satellites with $g-r\la 0.7$.  While these are rare in the observations, they do occur.  This could be caused by two reasons: (1) the strong AGN feedback quenches all the satellite galaxies instead of only most of them; (2) the low resolution means that gas stripping is too efficient, which quenches satellite galaxies progenitors prior to entering into the cluster \Romeel{or too quickly within the cluster}.  In the future we plan to run higher resolution \theth\ runs to check whether such resolution effects can impact these comparisons.

\section{Conclusions and discussions}
\label{sec:conc}

In this paper, we introduce a new suite of simulated galaxy clusters in \theth\ project called \simba, run using a feedback model based on the successful {\sc Simba} simulation~\citep{Dave2019}. We re-tune its feedback model parameters (mostly due to the coarse resolution of \theth\ clusters) by calibrating them with the observed stellar properties using one random cluster, specifically the stellar fraction, the BCG mass--halo mass relation and the satellite stellar mass function. Using this calibration we run all 324 clusters, and make detailed comparisons to both observational results and the \gadgetx\ runs which is also successful in reproducing many galaxy cluster properties and relations. By comparing the two hydrodynamic in \theth\, we aim to understand the origin of observational properties of galaxy clusters under different galaxy formation models.  We note that a comparison between \gadgetx\ and other galaxy formation models in \theth\ was done in \citet{Cui2018b}. Our main results are as follows:
\begin{itemize}
    \item Similar stellar mass fractions $f_*(<R_{500})$ are found at $z=0$ between \gadgetx\ and \simba, which are also in agreement with observational results. However, their gas fractions at $z=0$ are different ($\sim 5$ per cent in absolute mass fraction) at $M_{500}\sim 10^{14}\Msun$, with agreement for the most massive haloes and larger differences for low mass haloes. This difference is seen at all redshifts, either at a fixed halo mass or by tracking progenitors. This could be rooted mainly in the feedback scheme, of which \simba's kinetic feedback is very strong and efficiently blows gas out of haloes. More accurate gas fractions from observation are required to distinguish between models. 
    \item Though the stellar mass fraction is in agreement at $z=0$ in both models, \simba\ has a much stronger evolution of this fraction versus redshift than \gadgetx, especially at high redshift $z \gtrsim 1.5$. This could be caused by two reasons: (1) the star formation model is more efficient; (2) the quenching of galaxies is very quick (at mostly high redshift) and thorough. The high star formation rate in \simba\ produces a higher stellar mass fraction at high redshift, while at later epochs the complete shutdown of star formation makes its stellar mass fraction similar to \gadgetx\ by $z=0$.
    
    \item Using BCG stellar masses within fixed apertures, both \gadgetx\ and \simba\ are in agreement with observations from \cite{Kravtsov2018}. However, the BCGs (identified by the 6D \caesar\ galaxy finder) from \gadgetx\ have much brighter and bluer colour compared to \simba; observations agree better with \simba. This is corroborated by the mean BCG ages in \gadgetx\ being systematically lower than in \simba, with a long tail of younger BCGs.
    
    \item To test \simba's black hole accretion and feedback models, we compare three BH scaling relations from \simba\ versus $M_*$, $\sigma_*$, and $M_{500}$. \simba\ matches observations generally well, although it predicts higher black hole masses in the most massive galaxies, a shallower slope for the $M_\bullet-\sigma_*$ at the low-$\sigma_*$ end, and a shallower $M_\bullet-M_{500}$ relation at the more massive end.  A more careful comparison versus these data mimicking aperture and selection effects is required to assess the significance of these discrepancies.
    
    \item For satellite galaxies, we focus on satellite stellar mass function and colour-magnitude diagram. Although both agree with data for more massive galaxies, \simba\ shows better agreement to observation than \gadgetx\ at the lower mass end, which is only partly a result of tuning. The satellite galaxies in \simba\ agree very well with the $g-r$ vs. $r$ colour-magnitude diagram of the SDSS galaxies, while \gadgetx\ satellites are somewhat bluer. However, \simba\ shows less scatter than observations, with no satellites bluer than $g-r\la 0.7$ whereas observations indicate some.
\end{itemize}

The gas scaling relations are compared in the appendixes. As a result of \simba's strong feedback, its temperature--mass ($T-M$) relation tends to be a little bit higher than \gadgetx. However, in combination with its slightly lower gas fractions, the resulting integrated Sunyaev-Zeldovich decrement vs. mass ($Y-M$) relation is quite similar to that in \gadgetx.

The \simba\ runs in \theth\ provide a state of the art galaxy formation model with which to examine the evolution of galaxies, gas, and black holes within the largest virialized structures in the universe.  Many follow-up works using this suite are already underway.  For instance, for the BCG and satellite galaxies, we don't provide details on their evolution in this paper because (1) to disentangle BCG with ICL needs a careful study, which we will present in a companion paper (Ca\~nas et al. 2022); (2) we need to do a careful tracking to fully look into the the evolution history of satellite galaxies in hydro-simulations because as discussed in \cite{Behroozi2015}, halo and galaxy tracking is complicated within the cluster environment to properly account for mergers and tidal stripping \citep[see][and references therein, for the formation of the ultra-diffused galaxies without dark matter]{Ogiya2022}.  These sorts of studies can shed further light on how galaxies evolve within dense environments.

\section*{Acknowledgements}
We thank the referee Prof Peter Thomas for his valuable comments and suggestions on this paper. We would like to thank \theth\ collaboration for all the help on this project and this paper.

As part of \theth\ project, this work has received financial support from the European Union’s Horizon 2020 Research and Innovation programme under the Marie Sklodowskaw-Curie grant agreement number 734374, the LACEGAL project

The simulations used in this paper have been performed in the MareNostrum Supercomputer at the Barcelona Supercomputing Center, thanks to CPU time granted by the Red Espa\~{n}ola de Supercomputac\'{i}on. The CosmoSim database used in this paper is a service by the Leibniz-Institute for Astrophysics Potsdam (AIP). The MultiDark database was developed in cooperation with the Spanish MultiDark Consolider Project CSD2009-00064. This work also made use of the Gravity Supercomputer at the Department of Astronomy, Shanghai Jiao Tong University.

WC and RD are supported by the Science and Technology Facilities Council (STFC) AGP Grant ST/V000594/1. WC and YZ acknowledge the science research grants from the China Manned Space Project with NO. CMS-CSST-2021-A01 and CMS-CSST-2021-B01. This work is also supported by WC's Atracci\'{o}n de Talento Contract no. 2020-T1/TIC-19882 granted by the Comunidad de Madrid in Spain. further acknowledge the 
AK, GY, DC, RC and AC are supported by the Ministerio de Ciencia, Innovaci\'{o}n y Universidades (MICIU/FEDER) under research grant PGC2018-094975-C21. AK further thanks Lana Del Rey for video games.
DA is supported by the National Science Foundation Graduate Research Fellowship under Grant No. DGE 1746045.
MDP acknowledges support from Sapienza Universit\`a di Roma thanks to Progetti di Ricerca Medi 2019, RM11916B7540DD8D.
SE acknowledges financial contribution from the contracts ASI-INAF Athena 2019-27-HH.0, ``Attivit\`a di Studio per la comunit\`a scientifica di Astrofisica delle Alte Energie e Fisica Astroparticellare’' (Accordo Attuativo ASI-INAF n. 2017-14-H.0), INAF mainstream project 1.05.01.86.10, and from the European Union’s Horizon 2020 Programme under the AHEAD2020 project (grant agreement n. 871158).
YW is supported by NSFC grant No.11733010 and NSFC grant No.11803095.
XY is supported by the national science foundation of China (Nos. 11833005, 11890692) and by the China Manned Space Project with No. CMS-CSST-2021-A02.
DC is a Ramon-Cajal researcher.
KD acknowledges support by the Deutsche Forschungsgemeinschaft (DFG, German Research Foundation) under Germany’s Excellence Strategy - EXC-2094 - 390783311 and support through the COMPLEX project from the European Research Council (ERC) under the European Union’s Horizon 2020 research and innovation program grant agreement ERC-2019-AdG 882679.
RH acknowledges support from the STFC through a studentship.
UK acknowledges support from the STFC through grant number RA27PN.
YZ acknowledges the support by the National Science Foundation of China (11873038, 12173024).

%%%%%%%%%%%%%%%%%%%%%%%%%%%%%%%%%%%%%%%%%%%%%%%%%%
\section*{Data Availability}

The results shown in this work use data from \theth\ galaxy clusters sample. These data are available on request following the guidelines of \theth\ collaboration, at \url{https://www.the300-project.org}. The data specifically shown in this paper will be shared upon request to the corresponding author.
The plots and analyses done in this paper have extensively used these python packages: Ipython with its Jupyter notebook \citep{ipython}, astropy \citep{astropy:2018,astropy:2013}, NumPy \citep{NumPy} and SciPy \citep{Scipya,Scipyb}. All the figures in this paper are plotted using the python matplotlib package \citep{Matplotlib}.



\bibliographystyle{mnras}
\bibliography{paper,zotero_bcg} % if your bibtex file is called example.bib

%\begin{thebibliography}{99}
%\end{thebibliography}

%%%%%%%%%%%%%%%%%%%%%%%%%%%%%%%%%%%%%%%%%%%%%%%%%%

%%%%%%%%%%%%%%%%% APPENDICES %%%%%%%%%%%%%%%%%%%%%

\appendix

\section{The gas $T-M$ relation}
\label{app:1}

In this and the following appendices, we briefly check the gas properties from \simba\ by comparing the gas \Romeel{temperature--mass} relation ($T-M$; Appendix \ref{app:1}) and \Romeel{Sunyeav-Zeldovich (SZ) decrement--mass relation} ($Y-M$; Appendix \ref{app:2}) \citep[see][for a recent review]{Lovisari2021}. The $T-M$ relation will test \Romeel{whether} the gas temperature in \simba\ is affected by its strong feedback or not; while the $Y-M$ relation is checks whether the differences in gas fraction and temperature can be directly viewed from SZ observations.

\begin{figure}
	\includegraphics[width=0.5\textwidth]{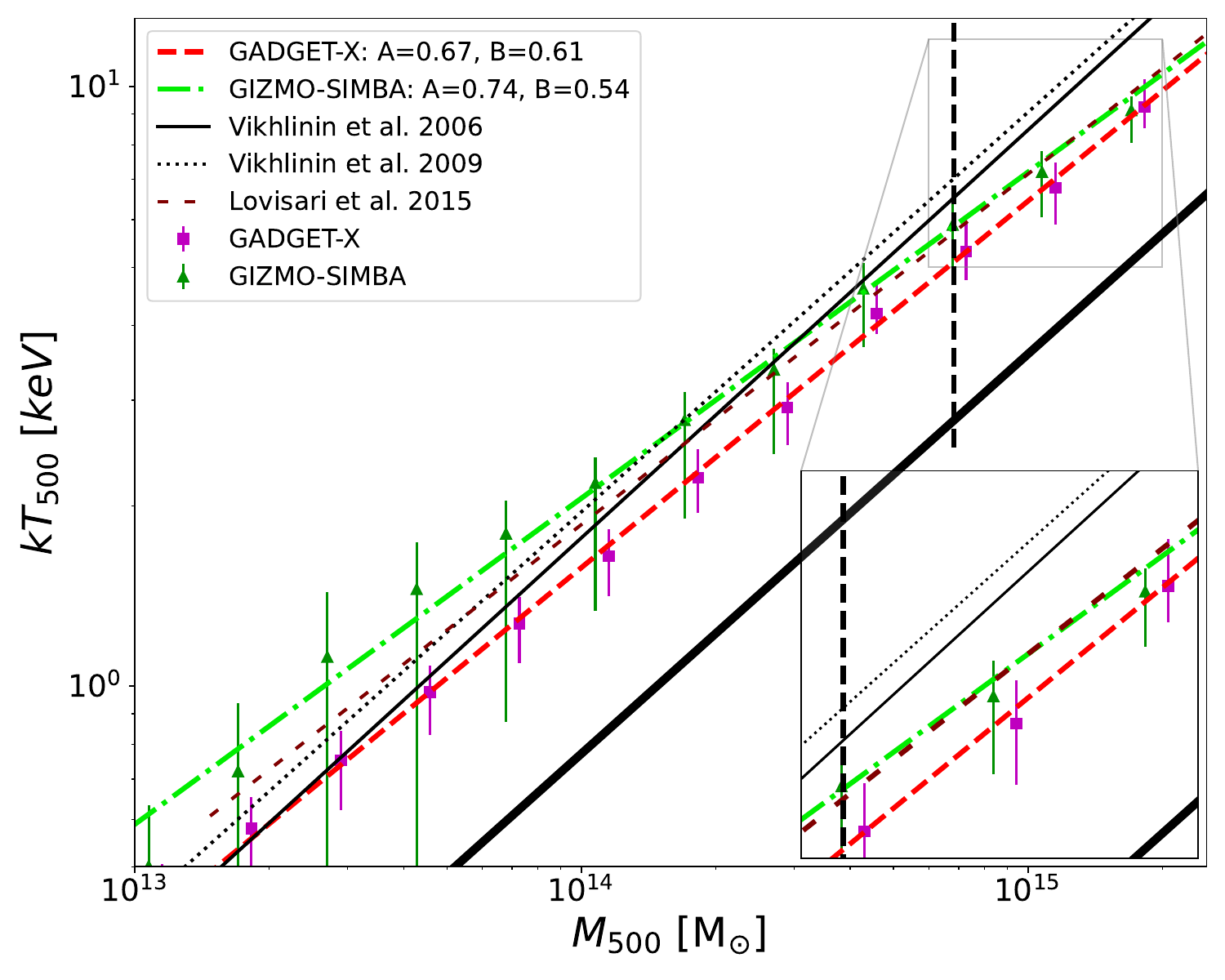}
    \caption{The gas mass-weighed temperature $T_{500}$ - halo mass $M_{500}$ relation. As indicated in the legend at top-left corner, data points from \gadgetx\ and \simba\ are shown with blue and green symbols with error bars (16th - 84th percentile) respectively. The solid and dotted black lines show the observational results from \citet{Vikhlinin2006} and \citet{Vikhlinin2009}, respectively. The maroon dashed line shows the fitting result from \citep[][]{Lovisari2015}. Our weighted fitting results from \gadgetx\ and \simba\ are presented by blue dotted and lime dashed lines, respectively. The thick solid black line shows the self-similar relation $\log_{10} T_{500} \propto 2/3 \log_{10} M_{500}$ is derived from self-similarity cluster approximations \citep[see][for example]{Bohringer2012}.}
    \label{fig:A1}
\end{figure}

Following \cite{300Cui2018}, we calculate the mass-weighted gas temperature within $R_{500}$ and add the scaling relation from \simba\ in \autoref{fig:A1}. \Romeel{In} the cluster mass range, \simba\ is in line with \gadgetx\ and the observational results, albeit a slightly higher temperature (see the zoomed-in \Romeel{inset} panel for details). This suggests that the gas heating inside these most massive clusters is reasonable. However, the median data points from \simba\ at the group halo mass range, $M_{500}<10^{14} \Msun$, seem to be \Romeel{slightly} higher than the other results, albeit \Romeel{with} a very large scatter. This \Romeel{is likely} related to the strong AGN feedback which is also efficiently expelling the gas out of $R_{500}$ as shown in \autoref{fig:1}. 

By applying a weighted fitting to $\log_{10} T_{500} = A + B*\log_{10} \frac{M_{500}}{6\times 10^{14} \Msun}$, in which the weight is given by the completeness of the halo \Romeel{counts}, \simba\ ($B=0.54$) seems \Romeel{to be} deviating from the self-similar slope much \Romeel{more} than \gadgetx\ ($B=0.61$). This is mainly because \Romeel{of} the high temperature in \simba\ at lower halo masses. Note that the median data points deviating from the fitting line at lower halo masses is mainly caused by the weighted fitting method. Furthermore, the slightly different fitting results between this work and \cite{300Cui2018} is because the new mass completeness based on \simba\ is adopted (see \cite{300Cui2022} for the changes). This indicates the fitting results are very sensitive to the weights. \Romeel{We note that observational results are quite uncertain within this mass regime, so it is too early to determine which model agrees better with data.}

\section{The gas Y-M relation}
\label{app:2}

\begin{figure}
	\includegraphics[width=0.5\textwidth]{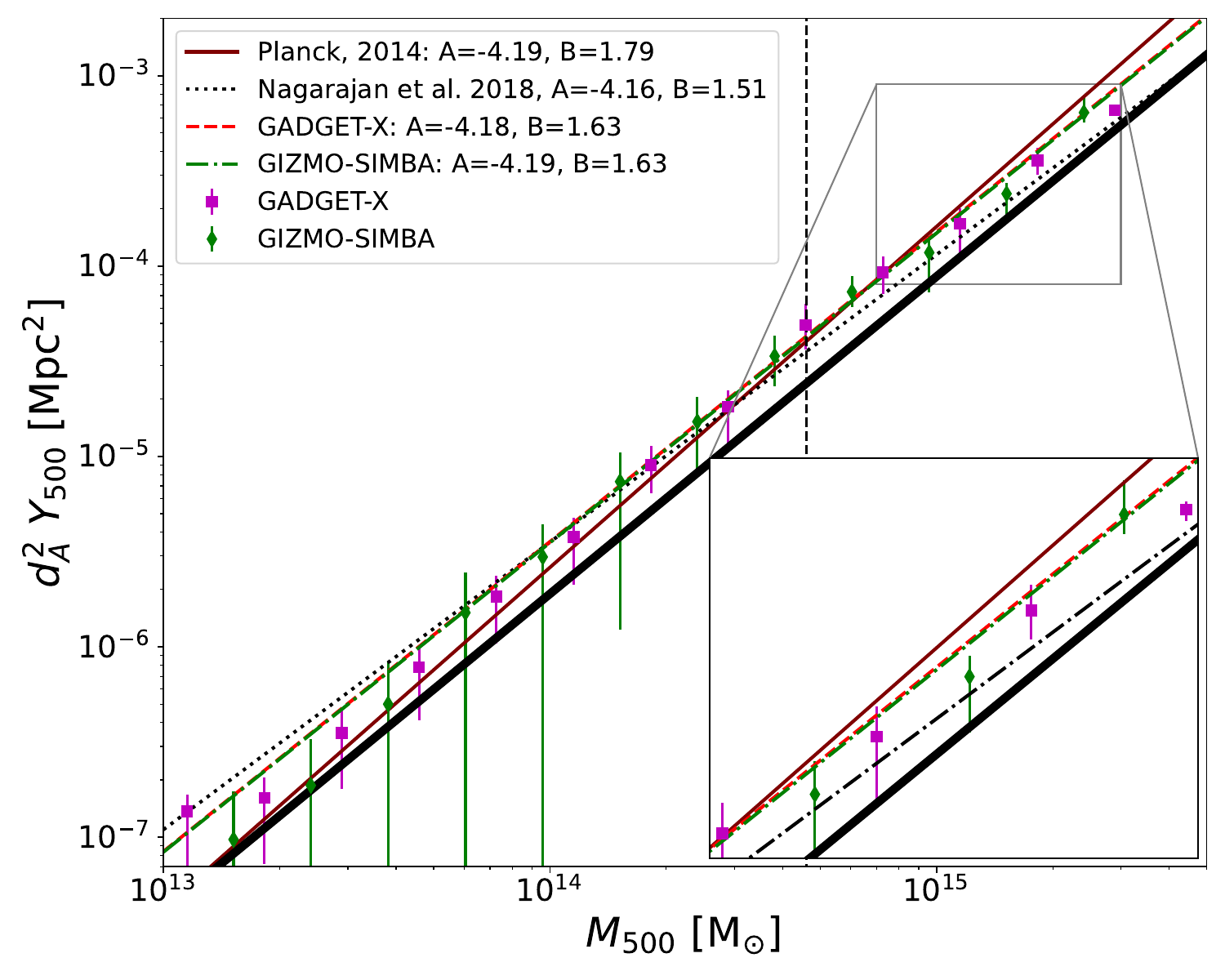}
    \caption{The updated Y-M relation from \simba. As indicated in the legend, two fitting results \citep{Planck2014,Nagarajan2018} from observations are shown by brown solid and black dotted lines, respectively. While the weighted fitting results with the parameters from the two hydro-simulations are shown by blue dashed and green dotted-dashed lines as before.}
    \label{fig:A2}
\end{figure}
%%%%%%%%%%%%%%%%%%%%%%%%%%%%%%%%%%%%%%%%%%%%%%%%%%
\Romeel{It is} also interesting to \Romeel{compare} the overall simulated SZ signal, the integrated SZ $Y_{500}$ value in this case, \Romeel{versus that} directly obtained from observation. \Romeel{For this}, we apply the {\sc PyMSZ} package\footnote{Publicly available at \url{https://github.com/weiguangcui/pymsz}, see \cite{Baldi2018} for the details of generating the kinetic SZ signal with this package.} to \simba\ simulated halos to generate mock $y-$maps along z-direction. Then, $Y_{500}$ \Romeel{is taken to be} the \Romeel{sum of the} $y$ values within $R_{500}$. Note that the ling-of-sight depth is set to $2\times R_{500}$ to be consistent with \Romeel{published results from} \gadgetx; we refer to \cite{Yang2022} for \Romeel{more a detailed examination of} projection effects. 

As shown in \autoref{fig:A2}, the median data points from \simba\ are comparable to the ones from \gadgetx. This is not surprising as the SZ$-y$ signal is proportional to gas density times temperature \citep[][and references therein]{Sunyaev1970}. The lower gas fraction in \simba\ seems to be compensated by its high temperature to give such an agreement. Note that the large scatters in gas mass fraction and temperature in \simba\ at the low halo mass range are also shown in this $Y_{500} - M_{500}$ relation. 
We also apply a weighted fitting of this $Y_{500} - M_{500}$ relation to $\log_{10} d_A^2 Y_{500} = A + B*\log_{10} \frac{M_{500}}{6\times 10^{14} \Msun}$ for the two hydro runs, where $d_A$ is the angular diameter distance. \Romeel{Given the similarity of the predictions,} it is not surprising to see such a good agreement in the fitting parameters between \simba\ and \gadgetx. Regarding the slope difference to the Planck result \citep[][]{Planck2014}, we note that the fitting line actually deviates from the median data points at both massive and low halo mass range: the slope from the median data points is more close to or even flatter than the self-similar slope of 5/3. Adopting the weighted fitting makes the slope dominated by these massive halos, and results in a value close to self-similar. Meanwhile at the low mass end, the slope from the data points is much steeper in both \gadgetx\ and \simba. If we do an unweighted fitting to these data points, we get a steeper slope which is comparable to the result from Planck. The different slopes for $Y_{500} - M_{500}$ relation across various mass ranges have been found in several works using different hydrodynamic simulations \citep[e.g.][]{Yang2022}. Therefore, \Romeel{using a fitting function with a single slope is not a good choice for representing the full mass range.}

\section{Resolution effects}
\label{app:3}
\begin{figure*}
	\includegraphics[width=\textwidth]{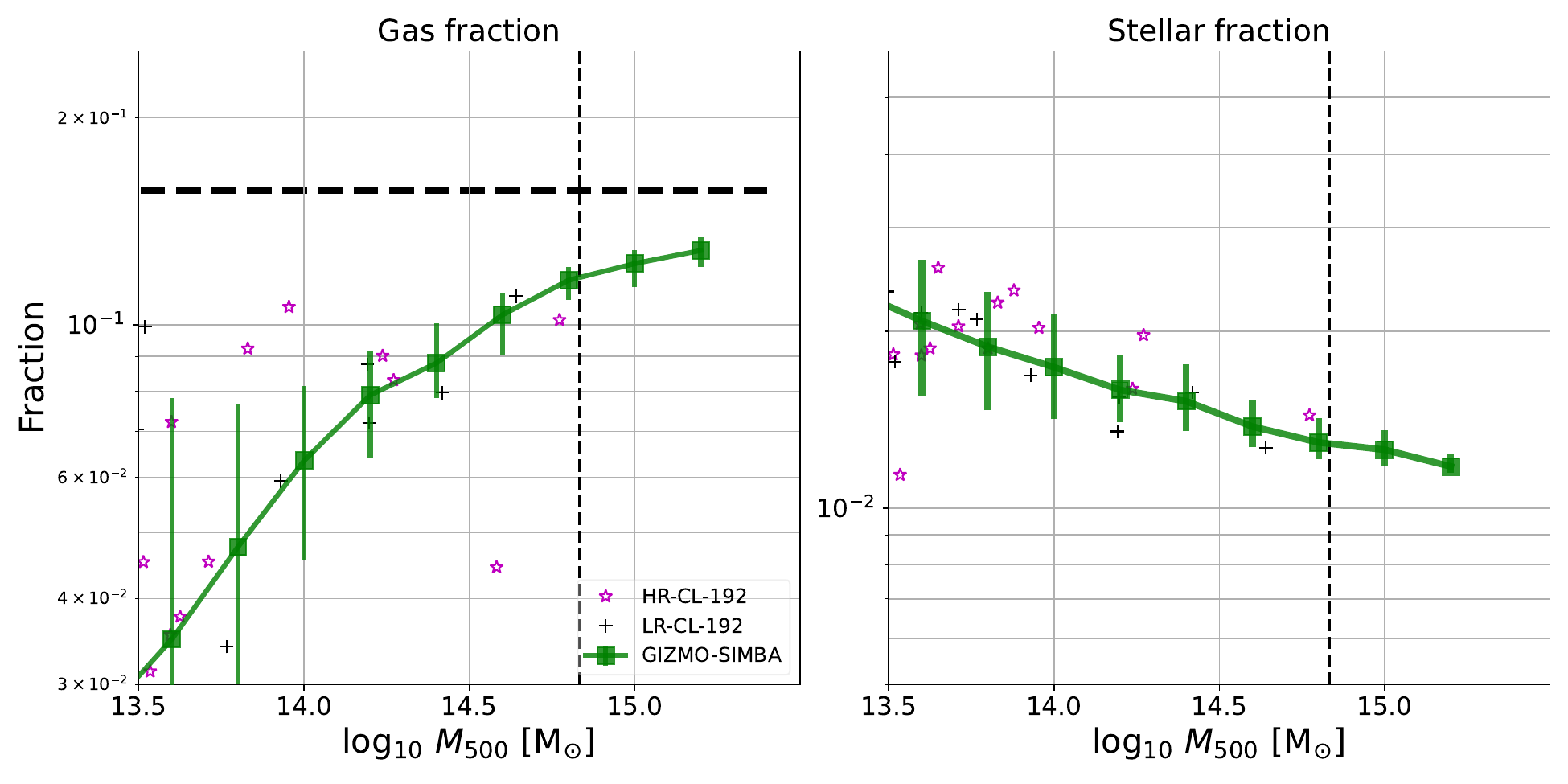}
    \caption{Similar to Fig.~\ref{fig:1}, baryon fraction within $R_{500}$. This plot focuses on comparing these fractions in the same zoomed-in region but one run with the default {\sc SIMBA} setup for a high-resolution IC, marked as `HR-CL-192' in open magenta stars, and one run in this paper with about 8 time lower resolution, marked as `LR-CL-192' with black cross. The overall statistical result from \simba\ , green solid line with error bars, in Fig.~\ref{fig:1} is also shown here for reference.  \Romeel{There is not a large systematic effect owing to resolution obvious for this single object, but a larger sample is needed to quantify the differences.}}
    \label{fig:A3}
\end{figure*}
As described in the main text, we re-tuned the baryonic feedback parameters for \theth\ galaxy clusters relative to {\sc Simba} due to their poor resolution.  \Romeel{This was because we found that} we cannot \Romeel{match key} observations using the default {\sc Simba} setup. It would be interesting to see whether the default {\sc Simba} setup works for the clusters which have a similar resolution as its original run. \Romeel{This requires doing a higher-resolution run of a \theth\ cluster, with a similar resolution to that in {\sc Simba}.}
We \Romeel{show the results of such a run with $8\times$ higher mass resolution} in this section using one of our cluster regions. 

In Fig.~\ref{fig:A3}, we show the comparisons for both gas and star fractions between the high-resolution run with our default run from the same region. It seems that these fractions are generally in agreement between the two runs\Romeel{; there are no obvious systematic differences, albeit with small numbers.  There may be a hint that gas fractions are slightly higher at low masses when run at higher resolution.  Furthermore,} we would caution that there maybe more significant differences in the detailed galaxy cluster properties, such as profiles \citep[as suggested in previous nIFTy comparison projects][]{Cui2016,Sembolini2016b,Elahi2016}. To statistically investigate and present these differences, we need a large sample of the high-resolution cluster runs, \Romeel{which is currently in progress}. 

% Don't change these lines
\bsp	% typesetting comment
\label{lastpage}
\end{document}